%% file: main.tex
\documentclass[conference]{IEEEtran}
\IEEEoverridecommandlockouts
\usepackage{booktabs} 
\usepackage[utf8]{inputenc} 
\usepackage[T1]{fontenc}    
\usepackage{hyperref}       
\usepackage{url}            
\usepackage{booktabs}       
\usepackage{amsfonts}       
\usepackage{microtype}      
\usepackage{graphicx}
\usepackage{amsmath}
\usepackage{amsthm}
\usepackage{dsfont}
\usepackage[lined,ruled,linesnumbered,noend]{algorithm2e}
\usepackage{algpseudocode}
\usepackage{multirow}
\usepackage{adjustbox}
\usepackage{xcolor}
\usepackage{balance,bm}
\usepackage[numbers]{natbib}
\usepackage{wrapfig}
\usepackage{color,soul}
\usepackage[labelformat=simple]{subcaption}

\newtheorem{theorem}{Theorem}
\newtheorem{lemma}[theorem]{Lemma}
\newtheorem{corol}[theorem]{Corollary}
\newtheorem{claim}[theorem]{Claim}
\newtheorem{definition}{Definition}
\providecommand{\internalFlag}{1}
\ifnum\internalFlag=1
\newcommand{\pnote}[1]{{\color{red}{\bf [#1]\\}}}
\newcommand{\td}[1]{{\color{red} Thang: {#1}}}
\else
\newcommand{\pnote}[1]{{}}
\newcommand{\todo}[1]{}
\newcommand{\td}[1]{}
\fi

\newcommand{\RP}{{SaPHyRa}$_{bc}$}
\newcommand{\SSP}{{SaPHyRa}}
\newcommand{\ABRA}{{ABRA}}
\newcommand{\KADBRA}{{KADABRA}}
\newcommand{\CSP}{\textsf{ISP}}
\newcommand{\PCSP}{\textsf{PISP}}
\newcommand{\SP}{\textsf{SP}}
\newcommand{\Hprob}{\textsf{HR}}
\newcommand{\exact}{\textsf{Exact}}
\newcommand{\exactbc}{\exact$_{bc}$}
\newcommand{\sample}{\textsf{Gen}}
\newcommand{\samplebc}{\sample$_{bc}$}

\newcommand{\sX}{\ensuremath{\mathcal{X}}}

\newcommand{\sC}{\ensuremath{\mathbf{C}}}
\newcommand{\sx}{\ensuremath{x}}
\newcommand{\VD}{\ensuremath{VD}}
\newcommand{\sY}{\ensuremath{\mathcal{Y}}}

\newcommand{\sH}{\ensuremath{\mathcal{H}}}
\newcommand{\sh}{\ensuremath{h}}
\newcommand{\dis}{\ensuremath{\mathcal{D}}}
\newcommand{\sD}{\ensuremath{{p}}}
\newcommand{\lf}{\ensuremath{f}}
\newcommand{\risk}{\ensuremath{\mathcal{R}}}
\newcommand{\riskest}{\ensuremath{\tilde{\mathcal{R}}}}
\newcommand{\riskemp}{\ensuremath{\mathcal{R}_\mathsf{e}}}
\newcommand{\loss}{\ensuremath{L}}
\newcommand{\erisk}{\ensuremath{\ell}}
\newcommand{\eRisk}{\ensuremath{L}}
\newcommand{\defeq}{\overset{\text{\tiny def}}{=}}
\newcommand{\weighte}{\ensuremath{\lambda}}

\def\BibTeX{{\rm B\kern-.05em{\sc i\kern-.025em b}\kern-.08em
    T\kern-.1667em\lower.7ex\hbox{E}\kern-.125emX}}
\begin{document}

\title{SaPHyRa: A Learning Theory Approach to\\ Ranking Nodes in Large Networks }

\author{
\IEEEauthorblockN{Phuc Thai}
\IEEEauthorblockA{\textit{Virginia Commonwealth University} \\
thaipd@vcu.edu}
\and
\IEEEauthorblockN{My T. Thai}
\IEEEauthorblockA{\textit{University of Florida} \\
mythai@cise.ufl.edu}
\and
\IEEEauthorblockN{Tam Vu}
\IEEEauthorblockA{\textit{Oxford University} \\
tam.vu@cs.ox.ac.uk}
\and
\IEEEauthorblockN{Thang Dinh}
\IEEEauthorblockA{\textit{Virginia Commonwealth University} \\
tndinh@vcu.edu}
}

\maketitle

\begin{abstract}
Ranking nodes based on their centrality stands a fundamental, yet, challenging problem in large-scale networks. Approximate methods can quickly estimate nodes' centrality and identify the most central nodes, but the ranking for the majority of remaining nodes may be meaningless. For example, ranking for less-known websites in search queries is known to be noisy and unstable.

To this end, we investigate a new node ranking problem with two important distinctions: a) \emph{ranking quality}, rather than the centrality estimation quality, as the primary objective; and b) \emph{ranking only nodes of interest}, e.g., websites that matched search criteria. We propose \underline{Sa}mple space \underline{P}artitioning \underline{Hy}pothesis \underline{Ra}nking, or \SSP, that transforms node ranking into a hypothesis ranking in machine learning. This transformation maps nodes' centrality to the expected risks of hypotheses, opening doors for theoretical machine learning (ML) tools. The key of \SSP{} is to  partition the sample space into exact and approximate subspaces. The exact subspace  contains samples related to the nodes of interest, increasing both estimation and ranking qualities. The approximate space can be efficiently sampled  with  ML-based techniques to provide theoretical guarantees on the estimation error.  Lastly, we present \RP, an illustration of \SSP{} on ranking nodes' \emph{betweenness centrality} (BC). By combining a novel bi-component sampling, a 2-hop sample partitioning, and improved bounds on the Vapnik–Chervonenkis dimension, \RP{} can effectively rank any node subset in BC. Its performance is up to 200x faster than state-of-the-art methods in approximating BC, while its rank correlation to the ground truth is improved by multifold. 
\end{abstract}


\input{Sections/1_Introduction}

\input{Sections/2_Preliminaries}

\input{Sections/3_Decomposition}
\input{Sections/3_HAP}
\input{Sections/3_SSP}
\input{Sections/4_BC}

\input{Sections/5_BCSubSet}
\input{Sections/6_Experiments}

\medskip
\noindent \textbf{Acknowledgment.} The work of My T. Thai is partially supported by NSF under award number CNS-1814614.
\bibliographystyle{IEEEtranN}
\bibliography{bc}
\balance

\end{document}

%% file: Sections/1_Introduction.tex
\vspace{-0.5cm}
\section{Introduction}
\label{sec::intro}
\vspace{-0.2cm}
Ranking nodes in a network poses a fundamental problem in network analysis, resulting in  various  centrality measures from degree centrality \cite{Newman10}, closeness centrality \cite{okamoto2008ranking}, betweenness centrality \cite{Bader07}, to Google’s PageRank \cite{bianchini2005inside}. It finds applications in identifying influential users in social networks, analyzing location in urban networks,  characterizing brain networks \cite{Newman10,Zhao15}, and so on.

In large networks the exact computation of centrality is intractable, thus, approximate methods \cite{okamoto2008ranking,Bader07,riondato2018abra,nathan2017graph} have been developed to quickly estimate nodes' centrality. Those methods are effective in identifying the most central nodes, however, the induced ranking for the majority of remaining nodes is often inaccurate. For nodes with small centrality values, a small error in estimation can result in a large perturbation in ranking, as seen in the ranking for less-known websites in search results \cite{ghoshal2011ranking}. The same challenge in ranking also arises from analyzing locations of lesser centrality in urban networks in which a majority of nodes have small betweenness centrality \cite{kirkley2018betweenness}. Thus, there is a \emph{lack of approximate methods that provide an accurate ranking of nodes in large-scale networks}. 

Moreover, most approximate methods to estimate centrality often produce the estimation for all nodes in the network \cite{Brandes07,Brandes08,Bader07,Borassi16}, even when only ranking for a small subset of nodes is required.  This often leads to the analysis of separate subnetworks, cut-off from a large network \cite{kirkley2018betweenness}, risking inaccurate assessment of nodes centrality in the complete network. \emph{Is it possible to design methods to rank node subsets substantially faster than ranking all nodes in the network?}

To this end, we investigate a new node ranking problem, called subset ranking, with two important distinctions: a) \emph{ranking quality}, rather than the estimation quality, as the primary objective; and b) \emph{ranking only target nodes}, e.g., websites matched search criteria. 

Our proposed solution is a framework, called Sample space Partitioning Hypotheses Ranking, or \SSP{}, that transforms ranking nodes into ranking hypotheses by mapping nodes' centrality to the expected risks of hypotheses, following the empirical risk minimization framework in machine learning \cite{Maurer09}. In \SSP,  we partition the sample space into exact and approximate subspaces and combine the evaluations of the hypothesis in both subspaces. The exact space contains  samples that are directly linked to the target nodes, providing  close ``heuristic'' estimation of the nodes' centrality. Further, the risks of the hypothesis in the approximate spaces can be effectively estimated using \emph{statistical learning theory} tools, including  Vapnik–Chervonenkis (VC) dimension \cite{shalev2014understanding} and empirical Bernstein \cite{Maurer09}.  

Lastly, we demonstrate the proposed framework on the task of ranking nodes' using betweenness centrality. Given a graph $G=(V, E)$, betweenness centrality (BC) \cite{Freeman77,Anthonisse71} defines the importance of each node $u \in V$ through the fraction of all-pairs shortest paths passing through $u$. Methods to approximate BC can be divided into two large groups: heuristics to either relax the shortest paths \cite{Everett05, Pfeffer12} or estimate BC via a subset of (non-uniformly) selected shortest paths \cite{Brandes07,Khopkar16, Chehreghani14} and sampling-based methods with additive error guarantees \cite{Brandes07,Brandes08,Bader07,Borassi16}. Unfortunately, the recent benchmark \cite{AlGhamdi17} using   96,000 CPUs and roughly 400 TB RAM indicates that existing methods either do not scale to large networks, e.g., Orkut network with 125 million edges, or produce \emph{poor ranking quality}. 

In contrast, our proposed solution, called \RP, offers both substantially better ranking quality and scalability. First, \RP{} has a new sampling method, called bi-component sampling. Inspired by the approach to compute exact BC in \cite{sariyuce2013shattering}, our new sampling limits the attention to only the shortest paths with both ends belonging to the same bi-component, thus, reduces the complexity of the sample space. Second, \RP{} deploys a 2-hop-based sample partitioning to guarantee non-zero estimation for nodes with small centrality. Third, \RP{} has a smaller sample complexity by reducing the VC dimension from $O(\log VD(G)$ in \cite{riondato2018abra} to $O(1)$ in many scenarios. Our experiments on large networks show that \RP{} is up to 200x faster than state-of-the-art methods in approximating BC, while its rank quality is improved by multifold. 


Our contributions are summarized as follows:
\vspace{-0.1cm}
\begin{itemize} 
\item We propose a novel formulation of ranking a node subset in large networks with the focus on the ranking quality and time-saving in ranking subset (but not all nodes). We also propose \SSP, a general framework to effectively rank nodes, especially, when nodes have small centrality values. \SSP{} provides an $(\epsilon, \delta)$-estimation for the nodes of interest, using fewer samples, yet, with higher ranking quality, thanks to its sample space partitioning strategy. 
\item We propose \RP, an illustration of \SSP{} for ranking nodes using betweenness centrality.  \RP{}  significantly improved ranking quality in comparisons to the state-of-the-art BC approximate methods. It also provides new VC-dimension bounds, i.e., tighter sample complexity.
\item We perform comprehensive experiments on both real-world and synthesis networks with sizes up to 100 million nodes and 2 billion edges.  Our experiments indicate the superior of \RP{}  algorithm in terms of both accuracy and running time in comparison to the state-of-the-art algorithms. 
\end{itemize} 
\vspace{-0.1cm}
\noindent
\textbf{Organization.} The rest of the paper is organized as follows: In section 2, we introduce the Ranking Subset and the Hypotheses Ranking problems. We propose \SSP{} framework to solve Hypotheses Ranking problem in section 3. To rank nodes using betweenness centrality, we develop \RP{} algorithm in section 4. In section 5, we present empirical evidence on the efficacy of \RP{} algorithm (and the proposed \SSP{} framework).

\noindent
\textbf{Related work.}
A few methods focus on estimating the rank without first computing the exact values of the centrality.
In \cite{saxena2017faster,saxena2016estimating}, Saxena et al. estimate the rank of nodes based on their closeness centrality. In \cite{wehmuth2013daccer,cabral2015mdaccer,steinert2017longitudinal}, heuristics are proposed to compute centrality based only on localized information restricted to a limited neighborhood around each node. Several recent works \cite{de2020approximating,grando2018machine,grando2018computing,kumar2015neural}  aim to approximate node centrality for large networks using neural networks and graph embedding techniques. Notably,
\cite{avelar2019multitask} demonstrate the multitask learning capability of the model to allow the ability to learn multiple centralities in the same model. While these heuristics can quickly estimate the nodes' ranking, there are no guarantees on the estimation errors. In contrast, we aim for fast ranking estimation with theoretical bounds on the estimation error.

Exact computation of betweenness centrality takes $O(MN)$ times in unweighted networks \cite{Brandes01}. Bader et al. \cite{Bader07} introduce an adaptive sampling algorithm to reduce the number of single-source shortest paths. Riondato et al. introduce a different sampling method that samples node pairs, resulting in faster computation with the same probabilistic guarantees on the estimation quality. In an algorithm called ABRA, The sampling method is further fine-tuned using Rademacher averages in \cite{riondato2018abra}. KADABRA is proposed by Borassi et al. \cite{Borassi16} to speed up the sample generation via a new Bread-first-search approach. These approaches provide approximate betweenness centrality for all nodes in the network with rigorous theoretical guarantees on the additive estimation errors. Unfortunately, the estimated centrality values result in poor ranking as shown in our experiments. Further, there is little saving in computational effort to estimate the centrality values for just a few nodes, compared to that for all the nodes in the network. 

Many advances in developing parallel and distributed algorithms to compute and estimate centrality in networks \cite{Bader07,sariyuce2013betweenness,mclaughlin2014scalable, bernaschi2018multilevel,van2020scaling}. We note that our effort in ranking nodes here is orthogonal to these efforts. Our sampling framework can be potentially combined with parallel and distributed methods to boost scalability.



%% file: Sections/2_Preliminaries.tex
\vspace{-0.05cm}
\section{Preliminaries}
\vspace{-0.05cm}
Consider a network, abstracted as a graph $G = (V,E)$ with $n=|V|$ nodes and $m= |E|$ edges. In our \emph{ranking subset problem}, we wish to rank the nodes in a subset $A \subseteq V$  according to a centrality measure 
 $c(.)$.  We focus on the case that it is computationally intractable to compute the exact centrality values $c(v)$ for $v \in A$, e.g., when the network has billions of edges or nodes. 
However, we should be able to estimate the centrality measure with some guaranteed error through sampling.




\vspace{-0.03cm}
\subsection{Ranking subset problem (RSP)}
\vspace{-0.01cm}
Let $A \subseteq V$ be a subset of nodes, called \emph{target nodes}, that we wish to rank using some centrality measure $c(.)$. 
Our goal is to produce a ranking of the nodes in $A$ that is close to the ground truth ranking. 


To produce the ranking of the nodes, we estimate the centrality values based on a set  of $N$ samples $x_1,\cdots,x_N$ and a function $g(\cdot,\cdot)$. For each node $v \in A$, we compute the approximation 
\vspace{-0.3cm}
$$\tilde{c}(v) = \frac{1}{N} \sum_{i=1}^n g(v,x_i).$$ 
\vspace{-0.3cm}

For example, consider the betweenness centrality that measures the importance of nodes in a network based on the fraction of shortest paths that pass through them. To estimate the betweenness centrality \cite{riondato2018abra},
we generate each sample $x_i$ ($i \in [N]$) as a random shortest path. More precisely, we first randomly select a pair of nodes $(u,v)$ in $V$. Then, we randomly select a shortest path $p$ between $u$, $v$ and set $x_i = p$. 
The function $g(v,x_i)$ is a binary function that outputs $1$ if $v$ is an inner node of $x_i$.
 
Another example is k-path centrality \cite{alahakoon2011k}. The k-path centrality measures the importance of nodes based on the fraction of $k$-hop paths that pass through them.
In k-path centrality estimation \cite{alahakoon2011k}, each sample $x_i$ ($i \in [N]$) is a random path that consists of at most $k$ edges. Here, we first randomly select a node $u$. Then, we perform an $l$-hop (where $l \le k$) random walk from $u$ and set $x_i$ as the $l$-hop random walk. 
The function $g(v,x_i)$ is a binary function that outputs $1$ if $v$ belongs to $x_i$.

We measure the quality of an estimation based on the \emph{ranking quality} and the \emph{estimation quality}.

\paragraph{Ranking quality}
To measure the ranking quality, we adopt Spearman's rank correlation \cite{spearman1904}. Other rank correlation measures such as Kendall's $\tau$ \cite{kendall1948rank} can also be used. 

Let $A=\{v_1, v_2,\ldots,v_{k}\}$ where $k = |A|$ is the size of $A$. Denote by $\mathbf{c}=<c(v_1), c(v_2), \ldots,c(v_k)>$ and $\mathbf{\tilde c}=<\tilde c(v_1), \tilde c(v_2), \ldots,\tilde c(v_k)>$ the centrality vector and the approximate centrality vector, respectively. 

As the ranks of the nodes are distinct integers between $1$ and $k=|A|$, the rank correlation can be computed using the following simple formula \cite{spearman1904}
\vspace{-0.3cm}
\begin{equation}
\label{eq:spearman}
	r_s = 1 - \frac{6 \sum_{i=1}^{k}dr_i^2}{k(k^2-1)},
	\vspace{-0.3cm}
\end{equation}
where $dr_i$ is the difference between the actual rank of $c(v_i)$ in $\mathbf{c}$ and the rank of its estimation $\tilde{c}(v_i)$ in $\tilde{ \mathbf{c}}$. 

\paragraph{Estimation quality}  We say $\mathbf{\tilde{c}}$ is an $(\epsilon, \delta)$-estimation of $\mathbf{c}$ if and only if
\vspace{-0.3cm}
\begin{equation}
    \Pr[\forall v_i \in A, |c(v)-\tilde{c}(v)| < \epsilon] \ge 1- \delta.
    \vspace{-0.3cm}
\end{equation}
To obtain an $(\epsilon, \delta)$-estimation of $\mathbf{c}$, it requires $O(\frac{1}{\epsilon^2}\left(\log n + \log \frac{1}{\delta})\right)$ number of sample on the full network and $O(\frac{1}{\epsilon^2}\left(\log |A| + \log \frac{1}{\delta})\right)$ on the subset $A$.

\subsection{RSP as a hypothesis ranking problem}
We provide the mapping from RSP to the hypotheses ranking problem, a fundamental problem in machine learning.  
\paragraph{Hypothesis ranking (\Hprob{}) problem}
Consider a (discrete) sample space $\sX$ and a distribution $\dis$ over $\sX$, where the probability of each sample $x \in \sX$ is denoted as
$\Pr_{\sx_0 \sim \dis}[\sx_0 = x]$. 
Each sample $x \in \sX$ is mapped with a label $y = \lf(x)$. Here, $\lf:\sX \rightarrow \sY$ labeling function. 
We will restrict the label space $\sY$ to 
be a two-element set, usually $\{0,1\}$ or $\{-1,+1\}$.
In a machine learning problem, 
the algorithm needs to learn a function $\sh: \sX \rightarrow \sY$, called hypothesis, which outputs a label $y$ for a sample $x$. 

We use a non-negative real-valued \emph{loss function} $\loss(y',y)$ to measure the difference between the output $y'$ of a hypothesis and the true label $y$. The \emph{expected risk} of a hypothesis $\sh$ is defined as the expectation of the loss function, i.e.,
\vspace{-0.3cm}
\begin{equation*}
	\risk_{}(\sh) \defeq \sum_{\sx \in \sX} \Pr_{\sx_0 \sim \dis}[\sx_0 = x]\loss\left(\sh(x),\lf(x)\right)
	\vspace{-0.3cm}
\end{equation*}


Given a set of $k$ hypotheses $\sH = \{\sh_1,\cdots,\sh_k\}$.  Our goal now is to rank the hypotheses based on the expected risk. That is to compute an $(\epsilon, \delta)$ approximation of the expected risks that can provide a high rank correlation to the expected risks.

\emph{Mapping from RSP to Hypothesis ranking.} The ranking subset problems in which there exist sampling-based method to estimate the centrality. In that case, we can design the hypotheses so that the centrality value $c(v_i)$ equals the expected risk of $h_i$. Ranking the nodes is now equivalent to ranking the hypotheses based on their expected risks.

\subsection{Ranking subset based on betweenness centrality (RSP$_{bc}$)}
Betweenness centrality (BC) of a node $v \in V$, denoted by $bc(v)$, measures the transitivity of $v$, i.e., how frequently $v$ lies on shortest paths among other nodes. Mathematically, define
\begin{equation}
bc(v) = \frac{1}{n(n-1)}\sum_{s \neq v \neq t \in V} \frac{\sigma_{st}(v)}{\sigma_{st}},
\end{equation}
 where $\sigma_{st}$ denotes the number of shortest paths from $s$ to $t$ and $\sigma_{st}(v)$ denotes the number of shortest paths from $s$ to $t$ that $v$ lies on, respectively.







Computing exact BC takes $O(mn)$ \cite{Brandes07}, where $n$ and $m$ are the number of nodes and edges in the graph, respectively. Thus, it is intractable for large networks. 

To approximate the BC, several sampling methods have been proposed, mapping the BC value of a node $v$ to the expectation of whether $v$ will lie on a random shortest path. By treating the probability that $v$ lies on a  random shortest path as an expected risk, we can turn the ranking node subset based on BC into a hypothesis ranking problem. 

\noindent \emph{Mapping from RSP$_{bc}$ to Hypothesis ranking.}
\label{subsubsec:bcsampling}
%
The \SP{} sample space consists of all shortest paths between two nodes in the graphs. To be precise, let $P_{st}$ be the set of all shortest paths from $s$ to $t$ in the graph. The sample space is defined as follows.
\begin{equation}
\label{eq:npsamplespace}
\sX_{b} = \{p|(s,t) \in V, p \in P_{st}\}.
\end{equation}




The \SP{} distribution $\dis_{b}$ is a distribution over the \SP{} sample space, where the probability of a shortest path $p$ from $s$ to $t$ is 
\begin{equation}
\label{eq:npdistribution}
	\Pr_{\sx\sim\dis_b}[x=p] = \frac{1}{n(n-1)} \frac{1}{\sigma_{st}}.
\end{equation}

For a shortest path $p$ from $s$ to $t$, we refer all nodes $v \in p$, except $s,t$, as \emph{inner nodes}.
We define a function $g_v:\sX_{bc} \rightarrow \{0,1\}$ as follows
\begin{equation}
\label{eq:gv}
	g(v,p) = \begin{cases}
		1 & \text{If } v \text{ is an inner node in } p\\
		0 & \text{Otherwise }\\
	\end{cases}
\end{equation}

Given the target nodes $A \subseteq V$, we define for each node $v \in A$ a hypothesis $h_v$ that maps each random path $p$ to a value $h_v(p) = g(v, p)$. We use 0-1 loss function and choose the labeling function $f$ that always output $0$ for any path $p$, i.e., $\loss(h_v(p),f(p)) = \mathds{1}_{h_v(p)= f(p)} = g(v, p)$.
Following \cite{riondato2018abra},  betweenness centrality of a node $v$ equals the expected risk $\risk_{}(\sh_v)$. 
\begin{lemma}
	For any node $v \in V$, we have
	\begin{equation*}
	\mathds{E}_{p \sim\dis_b}\loss(h_v(p),f(p)) =  \mathds{E}_{p \sim\dis_b}h_v(p) = bc(v) 
	\end{equation*}
\end{lemma}

%% file: Sections/3_SSP.tex
\section{\SSP: Sample space Partitioning Hypotheses Ranking}
In this section, we present  the sample space partitioning (\SSP{}) framework to solve the \emph{hypothesis ranking} (\Hprob{}) problem. 
An application of the \SSP{} framework in ranking  nodes in a subset using betweenness centrality will be presented later in Section \ref{sec:bc}. Note that, due to the space limitation, here, we omit the proofs of the lemmas. The detailed proofs are presented in the Appendix of the full version \cite{fullversion}.

\vspace{-0.3cm}
\subsection{Direct estimation.} 
\label{subsec:direct}
\vspace{-0.2cm}
An efficient solution is to estimate the expected risks 
through sampling and rank the hypotheses based on the estimation.
To be precise, consider a sequence of $N$ samples $\mathbf{x} = (\sx_1,\cdots,\sx_N)$
where, $x_i \sim \dis, \forall i \in [N]$, Here, we write $\sx \sim \dis$ to mean that $\sx$ is drawn from the distribution  $\dis$. 

The estimation of a hypothesis $\sh \in \sH$  over $X$ is computed as
\vspace{-0.3cm}
\begin{equation*}
\riskemp(\sh) \defeq \frac{1}{N}\left(\sum_{i=1}^N \loss(\sh(x_i),\lf(x_i))\right)
\vspace{-0.2cm}
\end{equation*}

\noindent\emph{Sampling complexity.} The goal here is to find the number of samples $N$ to ensure that with a probability of at least $1-\delta$ (where $\delta \in (0,1)$), the difference between the estimation and the expected risk of any hypothesis is smaller than $\epsilon \in (0,1)$, i.e., 
\vspace{-0.3cm}
\begin{equation}
\Pr \left[\forall h \in \sH, |\risk(h)-\riskemp(h)| < \epsilon\right] \ge 1-\delta
\label{eq:estimationguarantee}
\vspace{-0.2cm}
\end{equation}
In this case, we refer to this as an  \emph{$(\epsilon, \delta)$-estimation} of the expected risks. With $N =  O(\frac{1}{\epsilon^2}\left(\log |A| + \log \frac{1}{\delta})\right)$, we can guarantee an {$(\epsilon, \delta)$-estimation} of the expected risks. 
\noindent \textbf{Challenge in ranking hypotheses with low expected risks.} Ranking the hypotheses with low expected risk usually requires many more samples. Assume that we can only afford to generate enough samples to guarantee an $(\epsilon, \delta)$-estimation of the expected risks due to a time limit. Consider the two hypotheses that have the expected risks of $\mu_1, \mu_2$, respectively. If $\mu_1,\mu_2 < \epsilon$, i.e., $\epsilon > |\mu_1-\mu_2|$, there will be a high chance that the relative ranking of the two hypotheses is incorrect. 

\subsection{Sample space partitioning framework}
\label{subsec:ssp}
Our sample space partitioning (\SSP) framework partitions the sample space $\sX$ into two disjoint subspaces  $\sX = \hat{\sX} \bigcup \tilde{\sX}$ where $\hat{\sX}$ and $\tilde{\sX}$ are called \emph{exact subspace} and \emph{approximate subspace}, respectively.

The partition is done so that the exact subspace $\hat{\sX}$ will contain samples that are directly linked to the hypotheses in $\mathcal H$, aiding the estimation of the expected risks.  Especially, this will resolve the above challenge in estimating the hypotheses with low expected risks. The approximate subspace contains the majority of the samples, providing $(\epsilon,\delta)$-estimation guarantees for the expected risks. Combining them together, we have an estimation that provide both high ranking quality and theoretical guarantee on the estimation error.



\begin{algorithm}[ht]
	\SetKwInOut{Input}{Input}
	\SetKwInOut{Output}{Output}	
	\SetKwInOut{Param}{Parameters}	
	\SetKwInOut{Procedure}{Procedures}
	\Input{A sample space $\sX$, a  probability distribution $\dis$,
		 a labeling function $\lf$, 
		and a hypothesis class $\sH = \{\sh_1,\sh_2,\cdots,\sh_k\}$. Parameter $\epsilon,\delta \in (0,1)$}
	\Output{The rank of the hypotheses based on the expected risks}
	Partition $\sX = \hat{\sX} \cup \tilde{\sX}$;\\
	$\tilde{\dis}$ be the distribution over $\tilde{\sX}$ as in Eq. \ref{eq:disest} \\
	$(\hat{\lambda},\hat{\erisk}_1,\hat{\erisk}_2,\cdots,\hat{\erisk}_k) \gets $ \exact$(\cdot)$ \textcolor{gray}{ $\quad \quad \triangleright$ Compute the exact value in exact subspace $\hat{\sX}$}\\
	$\lambda \gets 1 - \hat{\lambda}$\\
	$\epsilon' = \epsilon / \weighte$\\
	$N_0 \gets \frac{c}{\epsilon'^2}\ln 1/\delta$   \textcolor{gray}{ $\quad \triangleright$ the initial \#samples}\\
	$N_{max} \gets \frac{c}{\epsilon'^2}(VC(\sH)+\ln 1/\delta)$  \textcolor{gray}{ $ \quad \triangleright$ the maximum \#samples} \\ 
	$N\gets N_0$ and $\mathbf{x} \gets \emptyset$ \\	
	Allocate the probability error $\delta_i$ such that Eq. \ref{eq:sumerror} is satisfied \\
	\For{$rd := 1$ to $\lceil\log(\frac{N_{max}}{N_0})\rceil$}
	{
		\For{$j := |\mathbf{x}+1|$ to $N$}{
			Genrate samples $x_j \sim \tilde{\dis}$ using \sample$(\cdot)$ and add $x_j$ to $\mathbf{x}$ 
		}
		\For{$i:=1$ to $k$}{
		$\mathbf{z}^{(i)} \gets [z_j^{(i)} = \loss(h_i(x_j),\lf(x_j))]_{j = 1..N}$\\ 
		$\epsilon_i \gets \varepsilon(N,\delta_i,Var(Z^{(i)}))$ \\ 
	}
		\If {$\max_{i\in[n]}\epsilon_i \leq \epsilon'$}{
			\textbf{Break}
		}
		$N = \min(2N, N_{\max})$
	}
	\For{$i:=1$ to $k$}{
		$\tilde{\erisk}_i \gets \frac{1}{N}\left(\sum_{j=1}^N\loss(\sh(x_j),\lf(x_j)\right)$\\
		${\erisk}_i \gets \hat{\erisk}_i +   \weighte\tilde{\erisk}_i $
	}
	
	Return $(\erisk_1,\erisk_2,\cdots,\erisk_k)$ and the rank of the hypotheses
	\caption{Sample Space Partitioning (\SSP{})}
	\label{alg::ssp}
\end{algorithm} 

Specifically, for each hypothesis $h_i$, we combine the expected risk  the expected risks of $h_i$ on the exact subspace $\hat{\erisk}_i$ and the estimation $ \tilde{\erisk}_i$ of the expected risks on the approximate subspace as follows 
\begin{equation}
\erisk_i = \hat{\erisk}_i +   \weighte \tilde{\erisk}_i,
\end{equation}
where $\lambda = \Pr_{x \sim \dis}[x \in \tilde{\sX}]$ be the probability that a random sample 
$x \sim \dis$ belongs to the approximate subspace.

\noindent
\textbf{Exact subspace.} We select the exact subspace $\hat{\sX}$ such that the expected risks on the exact subspace should provide good estimations for the hypotheses with small expected risks. At the same time, there should be an algorithm that can efficiently compute the expected risks on the exact subspace.
Let $\hat{\eRisk} = (\hat{\erisk}_1,\cdots,\hat{\erisk}_k)$ be the expected risks of $\sH = \{h_1,\cdots,h_k\}$ on exact subspace $\hat{\sX}$.
For each hypothesis $\sh_i$ ($i \in [k]$), $\hat{\erisk}_i$ is defined as follow,
\vspace{-0.1cm}
\begin{equation}
\hat{\erisk}_i \defeq \sum_{\sx \in \hat{\sX}} \Pr_{\sx_0 \sim \dis}[\sx_0 = x]\loss\left(\sh(x),\lf(x)\right)
\label{eq:exactrisk}
\vspace{-0.1cm}
\end{equation}
We assume the expected risks on the exact subspace can be computed via an algorithm \exact{} that returns
$(\hat{\lambda},\hat{\erisk}_1,\hat{\erisk}_2,\cdots,\hat{\erisk}_k)$, in which, for all $i \in [k]$, $\hat{\erisk}_i$ is computed as in Eq.~\ref{eq:exactrisk}.
An efficient partitioning of the sample space will need to provide close estimations for all the hypotheses in $\mathcal H$. 

%

\noindent
\textbf{Approximate subspace.} The approximate subspace $\tilde{\sX} = \sX \setminus \hat{\sX}$ contains the samples outside the exact subspace. We assume that there exists an effective algorithm to draw samples from the approximate subspace $\tilde{\sX}$. 
Let $\tilde{\dis}$ be a distribution over the approximate subspace  $\tilde{\sX}$, where the probability of a sample $\sx \in  \tilde{\sX}$ is
\vspace{-0.2cm}
\begin{equation}
\label{eq:disest}
\Pr_{\sx_0 \sim \tilde{\dis}}[\sx_0 = x] 
 = \frac{1}{\weighte }\Pr_{\sx_0 \sim \dis}[\sx_0 = x].
 \vspace{-0.2cm}
\end{equation}

For each hypothesis $h_i \in \sH$ ($i \in [k]$), we denote $\riskest(h_i)$ as the expected risk of $h$ on the distribution $\tilde{\dis}$, i.e.,
 \vspace{-0.2cm}
\begin{equation}
	\riskest(h_i) = 
	\sum_{x \in \tilde{\sX}} \Pr_{\sx_0 \sim \tilde{\dis}}[\sx_0 = x]\loss\left(\sh(x),\lf(x)\right).
	\label{eq:expectedrisk-estimation}
	 \vspace{-0.2cm}
\end{equation}
Let $\epsilon' = \epsilon/ \weighte$ and $(\tilde{\erisk}_1,\cdots,\tilde{\erisk}_n)$  be an $(\epsilon',\delta)$-estimation  of the expected risk on the distribution $\tilde{\dis}$, i.e., 
\vspace{-0.2cm}
\begin{equation}
\Pr \left[\forall i \in [k], |\riskest(h_i)-\tilde{\erisk}_i| < \epsilon'\right] \ge 1-\delta.
\label{eq:estimationrisk}
 \vspace{-0.2cm}
\end{equation}





\begin{lemma}
	\label{le:SSP}
	For any partition of $\sX = \hat{\sX} \cup \tilde{\sX}$.
	Let $(\hat{\erisk}_1,\cdots,\hat{\erisk}_n)$ be the expected risks  on exact subspace $\hat{\sX}$ (see Eq.\ref{eq:exactrisk}).
	Let $\riskest(h_i)$ be the expected risk of a hypothesis $h_i$ on the approximate subspace $\tilde{\sX}$ (see Eq.\ref{eq:expectedrisk-estimation}). Then, we have
 \vspace{-0.2cm}
	\begin{equation*}
	\forall i \in [k], \hat{\erisk}_i + \riskest(h_i) = \risk(h_i).
 \vspace{-0.2cm}
	\end{equation*}
	
\end{lemma}

\subsection{Risk Estimation in the Approximate Subspace} \label{subsec:estimation}
To estimate the risk within the approximate subspace we adopt two theoretical machine learning tools, namely, \emph{adaptive sampling} and \emph{Vapnik–Chervonenkis (VC) dimension}.

\noindent \textbf{Adaptive sampling.}
We apply empirical Bernstein' inequality \cite{Maurer09} to construct an \emph{adaptive sampling} method to estimate  expected risk on the approximate subspace.


 We allocate the error probability $\delta_i$ for each hypothesis $h_i$ such that
\begin{equation}
\label{eq:sumerror}
\sum_{i=1}^k 2\delta_i = \frac{\delta}{\lceil\log(\frac{N_{max}}{N_0})\rceil}.
 \vspace{-0.2cm}
\end{equation}
To minimize the number of iterations, we optimize the allocation of $\delta_i$ as follows. 
We first compute a sample variance $v_i$ for each hypothesis $h_i$ by taking $N_0 = \frac{c}{\epsilon'^2}\ln 1/\delta$  number of samples. Note that, the samples here are independent with the samples in $X$. 
For each hypothesis $h_i$, we use the sample variance $v_i$ and set $\epsilon_i = \epsilon$ for all $\epsilon_i$, to compute $\delta_i$ that satisfies Eq.\ref{eq::epsilon} (this can be done by binary search). 
After that, we rescale $\delta_i$ such that Eq. \ref{eq:sumerror} is satisfied.

Next, we compute the empirical risk. 
Initially, we have a list of $N_0$ samples. After each iteration, we double the number of samples and measure the current error probability using empirical Bernstein' inequality \cite{Maurer09}.

\begin{lemma}[Empirical Bernstein' inequality (Theorem 4 \cite{Maurer09}]
	\label{le:Bernstein}
	Let $\mathbf{z} = (z_1,z_2,\cdots,z_N)$ be a vector of independent identically random variables. Let $\mu$ be the expected value of a random variable $z_j$ ($j \in [N]$) and $\delta_0 \in (0,1)$.
	Then, we have,
	 \vspace{-0.2cm}
	\begin{equation}
	\Pr\Big[\mu- \frac{1}{N}\sum_{j=1}^N z_j \le \varepsilon(N,\delta_0,Var(\mathbf{z}))\Big] \ge 1-\delta_0,
 \vspace{-0.2cm}
	\end{equation}
	where 
	\vspace{-0.2cm}
	\begin{equation*}
	\varepsilon(N,\delta_0,Var(\mathbf{z})) = \sqrt[]{1/N\left(2Var\left(\mathbf{z}\right) ln\frac{2}{\delta_0}\right)} + {7ln\frac{2}{\delta_0}}/{3N}
	 \vspace{-0.2cm}
	\end{equation*}
	and $Var(\mathbf{z})$ is the sample variance 
	 \vspace{-0.2cm}
	\begin{equation*}
	Var(\mathbf{z}) = \frac{1}{N(N-1)}\sum_{1 \leq j_1 < j_2 \leq N }(z_{j_1} - z_{j_2})^2. \label{eq:var}
	 \vspace{-0.2cm}
	\end{equation*}
\end{lemma}

In Lemma \ref{le:Bernstein}, we only show one-sided error. We can show the errors on both sides by considering the random variables $z'_j = 1 - z_j$. 
Then, using union bound, we have
 \vspace{-0.2cm}
	\begin{equation*}
	\label{eq:bernstein}
\Pr\Big[\Big|\mu- 1/N\sum_{j=1}^N z_j\Big| \le \varepsilon(N,\delta_0,Var(\mathbf{z}))\Big] \ge 1-2\delta_0
 \vspace{-0.2cm}
\end{equation*}

Let $\mathbf{x} = (x_1,\cdots,x_N)$ be the current list of samples. 
For each hypothesis $h_i$, let 
$\mathbf{z}^{(i)} = (z^{(i)}_1\cdots,z^{(i)}_N)$ be the list of random variables for $h_i$, where 
 \vspace{-0.2cm}
\begin{equation*}
z^{(i)}_j = \loss(h_i(x_j),\lf(x_j))
 \vspace{-0.2cm}
\end{equation*}

We compute the error $\epsilon_i$ for each hypothesis $h_i$ base on $\delta_i$ and sample variance $V(\mathbf{z}^{(i)})$ as in Lemma \ref{le:Bernstein}, i.e., 
 \vspace{-0.2cm}
\begin{equation}
\label{eq::epsilon}
\epsilon_i = \varepsilon(N,\delta_i,Var(\mathbf{z}^{(i)}))
 \vspace{-0.2cm}
\end{equation}

If the maximum value of $\epsilon_i$ is smaller than or equal to $\epsilon'$, we stop the algorithm and take the current estimation of the expected risks.

\noindent {\textbf{Reducing sample complexity using VC-dimension.} }
We can reduce the number of samples
using {VC-dimension}, a standard complexity measure for concept classes in probably approximately correct (PAC) learning \cite{shalev2014understanding}. 

%

\begin{figure}[!ht]
	 \vspace{-0.2cm}
	\centering 
	\begin{subfigure}{0.36\textwidth}
		\vfill
		\includegraphics[width=\linewidth]{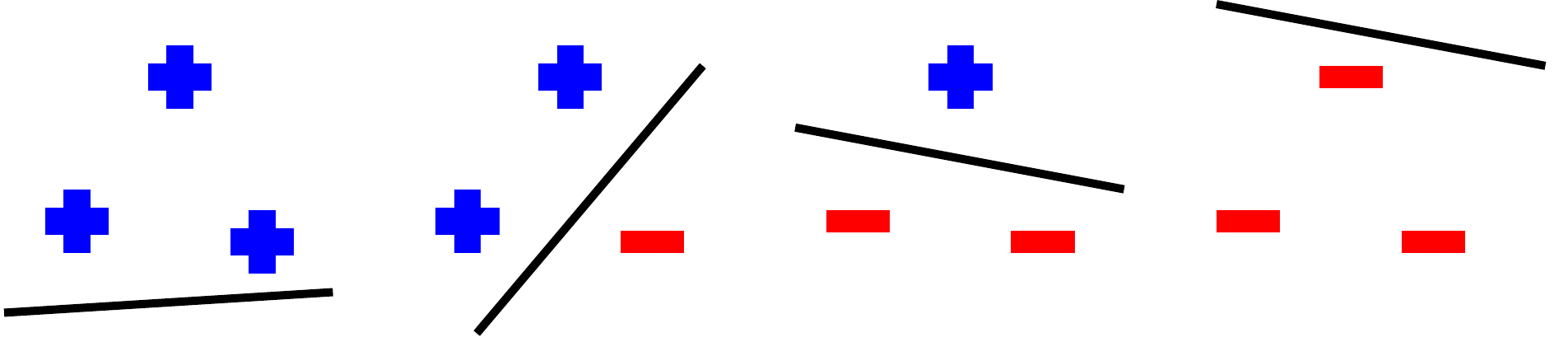}
		\caption{A set of 3 points that can be shattered. We can always find a hypothesis to separate positive data points from negative data points. \\ } 
		\label{fig:VC1}
		\vfill
	\end{subfigure}	
\hfill
	\begin{subfigure}[]{0.09\textwidth}
				\vfill
		\includegraphics[width=\linewidth]{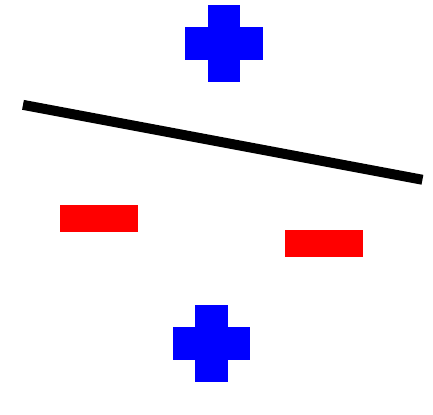}
		\caption{Any set of 4 points cannot be shattered.} 
					\label{fig:VC2}
				\vfill
	\end{subfigure}	
 \vspace{-0.2cm}
	\caption{\small The VC-dimension of the class $\sH$ of threshold functions.}

	\label{fig:VC}
	 \vspace{-0.2cm}
\end{figure}

Intuitively, VC-dimension measures  the capacity of a hypothesis class. It is defined as the cardinality of the largest set of points that the algorithm can shatter.
 \vspace{-0.2cm}
\begin{definition}[VC-dimension]
	For any subset $S = \{x_1,\cdots,x_w\} \subseteq \sX$, let $\sH_{S} = \{(\sh(x_1),\cdots,\sh(x_w)):\sh \in \sH\}$ be the restriction of $\sH$ to $S$. We say $\sH$ shatters $S$ if the restriction of $\sH$ to $S$ is the set of all functions from $S$ to $\{0,1\}$, i.e., $|\sH_{S}| = 2^w$.
	The VC-dimension of a hypothesis class $\sH$, denoted $VC(\sH)$ is the maximum size of a set $S \subseteq \sX$ that can be shattered by $\sH$. 
\end{definition}
 \vspace{-0.2cm}
For example, consider a two-dimensional binary classification problem where we need to label each point in a two-dimensional plane as positive (or $1$) or negative (or $0$). Here, each hypothesis $h \in \sh$ is a  straight line in which all data points above the line are labeled as positive and all data points below the line are labeled as negative. 
We say the hypothesis class $\sH$ shatters a set of points (samples) $S$ if for any labeling on S, we can always find a hypothesis to separate positive data points from negative data points.
We now show that the VC-dimension $VC(\sH) = 3$. 
Indeed, there exist a set of 3 points that can be shattered by $\sH$ (see Fig. \ref{fig:VC1}). 
Plus, by Radon's theorem \cite{bajmoczy1979common}, we can divide the four points into two subsets with intersecting convex hulls. Thus, it is not possible to separate the two subsets with a straight line (see Fig. \ref{fig:VC2}). 


We can bound the number of samples using VC-dimension as follows.
 \vspace{-0.2cm}
\begin{lemma}[THEOREM 6.8 in \cite{shalev2014understanding}]
	\label{le:VC}
	Given a hypothesis space $H$, defined over sample space $\sX$, such that $h_i: \sX \rightarrow \sY$ for $\forall h_i \in H$. $VC(\sH)$ is the VC dimension of $H$. Let $\mathbf{x} =  (\sx_1,\cdots,\sx_N)$ be a collection of independent identically distributed random events, taken from sample space $\sX$, selected by probability $\sD$. 
	We can obtain an $(\epsilon, \delta)$-estimation of the expected risk with 
	 \vspace{-0.2cm}
	\begin{equation}
	N = \frac{c}{\epsilon^2}(VC(\sH)+ln\frac{1}{\delta}),
	 \vspace{-0.2cm}
	\end{equation}
	where constant $c$ is approximately $0.5$. 
\end{lemma}

Thus, we can obtain an $(\epsilon, \delta)$-estimation of the expected risk with $N = \frac{c}{\epsilon^2}(SC(\sH)+ln\frac{1}{\delta})$ samples.

Now, we present a bound for the VC-dimension of a hypothesis class $\sH$.

\begin{lemma}
	For a sample $x \in \sX$, let $\pi(x)$ be the number of hypotheses $h \in \sH$ such that $h(x) = 1$.
	Let $\pi_{\max} = \max_{x\in\sX} \pi(x)$.
	 \vspace{-0.2cm}
	\begin{equation*}
	VC(\sH) \leq \lfloor \log(\pi_{\max}) \rfloor + 1.
	 \vspace{-0.2cm}
	\end{equation*}
	\label{lemma:vc}
\end{lemma}
\vspace{-0.5cm}

\subsection{Correctness and Complexity.}
We first state the correctness of our framework in providing theoretical guarantee on the estimation error.
\begin{theorem}
	\label{LEMMA:SSP}
	Consider a distribution $\dis$ over the sample space $\sX$, where the probability of a sample $\sx \in \sX$ is $\sD(\sx)$, a label space $\sY$, a label function $\lf$, and a hypothesis class $\sH = \{\sh_1,\cdots,\sh_k\}$. If Algorithm $\exact$ returns the expected risks on the exact subspace as in Eq. \ref{eq:exactrisk}, and Algorithm $\sample$ return a random sample $x \sim (\tilde{\sX},\sD)$. 
	Then,  $(\erisk_1,\erisk_2,\cdots,\erisk_k)$, returned by Algorithm \ref{alg::ssp}, is an 
	$(\epsilon,\delta)$-estimation of the expected risks
	 i.e.,
	 \vspace{-0.2cm}
		\begin{equation*}
	\Pr \left[\forall i \in [k], |\risk(h_i)-\erisk_i| < \epsilon\right] \ge 1-\delta.
	 \vspace{-0.2cm}
	\end{equation*}
\end{theorem}


%




The sample complexity is a function of the weight of the approximate space, i.e., $\lambda$ and the VC-dimension of the hypotheses $VC(\mathcal H)$.
\begin{lemma}
	\label{corol:worstcase}
	The worst-case number of samples in Algorithm \ref{alg::ssp} is 
	 \vspace{-0.2cm}
	\begin{equation}
	\frac{c\weighte^2}{\epsilon^2}(VC(\sH)+\ln 1/\delta),
	 \vspace{-0.2cm}
	\end{equation}
	which is reduced by a factor of $1/\lambda^2$, 
	in comparison with the direct estimation approach.
\end{lemma}

\noindent \emph{Variance reduction analysis.} We show that our \SSP{} framework results in random variables with generally smaller variances. Thus, the expected risks can be estimated with fewer samples.  From Eq. \ref{eq::epsilon}, if the variance is reduced by a factor of $\alpha$, we can also reduce the number of samples by a factor of approximately $\alpha$ 
to achieve the same error. Indeed, in Eq. \ref{eq::epsilon}, the first part is $\Theta(\sqrt{N})$ times bigger than the second part. 
Thus, if $N$ is big enough, we can approximate 
 \vspace{-0.2cm}
\begin{equation*}
N \approx \epsilon_i^2 \left(2Var\left(\mathbf{z}^{\left(i\right)}\right) ln\frac{2}{\delta_i}\right).
 \vspace{-0.2cm}
\end{equation*}

Let $Z^{(i)}$ be the random variable that represents the output of the loss function of $h_i$ on a sample that is drawn from the distribution $\tilde{\dis}$.
Let $Z'^{(i)}$ be the random variable that represents the output of the loss function of $h_i$ on a sample that is drawn from the distribution $\dis$. 
Let $\hat{\erisk}_i = \hat{\mu}_i$ be expected risk on the exact subspace of $h_i$. 
Let $Var(Z^{(i)})$, $Var(Z'^{(i)})$ be the variances of $Z^{(i)}$, $Z'^{(i)}$. 
The expected value of $Z'^{(i)}$ and ${Z}^{(i)}$ are $\mu_i$ and $\mu_i - \hat{\mu}_i$, respectively. 
Recall that, in this work, we consider $0$-$1$ loss function. Thus, $Z'^{(i)}$ and ${Z}^{(i)}$ Bernoulli  random variable.
Hence, we have, $Var(Z'^{(i)})	= {\mu}_i(1-{\mu}_i)$, and $Var({Z}^{(i)})	= (\mu_i-\hat{\mu}_i)(1-\mu_i+\hat{\mu}_i)$.
Therefore, the reduction in the variance is
\begin{equation*}
\frac{Var(Z^{(i)})}{Var(Z'^{(i)})} = \frac{{\mu}_i(1-{\mu}_i)}{(\mu_i-\hat{\mu}_i)(1-\mu_i+\hat{\mu}_i)} 
\end{equation*}

\begin{claim}
	\label{claim:variance}
	Consider a hypothesis $h_i \in \sH$ in which the expected risk $\mu_i < 1/2$. We have, 
	$Var(Z^{(i)}) < Var(Z'^{(i)}).$
	Specifically, if $\mu_i \ll 1$, we have
	 \vspace{-0.2cm}
	$$\frac{Var(Z^{(i)})}{Var(Z'^{(i)})}  \approx \frac{\mu_i}{\mu_i - \hat{\mu}_i}
	 \vspace{-0.2cm}$$
\end{claim}

In this work, we assume the expected risk $\mu_i$ of any hypothesis $h_i \in \sH$ is smaller than $1/2$. Here, $1/2$ is the expected risk of a baseline hypothesis that randomly output $0$ and $1$ with the same probability. In this case, $\frac{\mu_i}{\mu_i - \hat{\mu}_i} < 1$. From claim \ref{claim:variance}, the random variables in \SSP{} framework have smaller random variables. In other words, \SSP{} framework uses a smaller number of samples to achieve the same error guarantee.

%% file: Sections/4_BC.tex
 \vspace{-0.2cm}
\section{\RP: Ranking node subset with Betweenness centrality}
\label{sec:bc}
\begin{figure*}[!ht]
	\centering
	\begin{subfigure}{0.25\textwidth}
		\includegraphics[width=\linewidth]{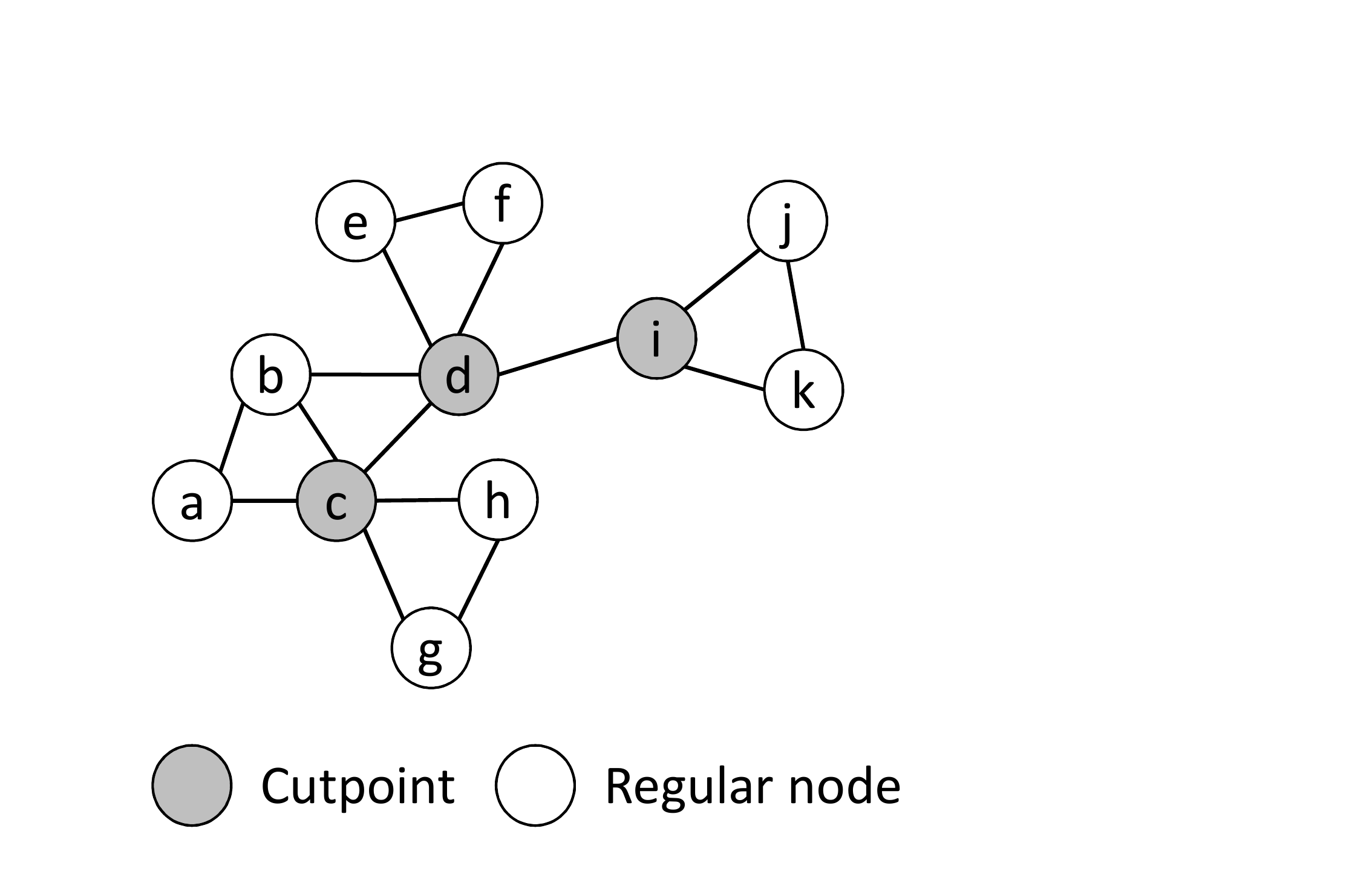}
		\caption{A graph $G$. 
		} 
	\end{subfigure}	
	\hfill
	\begin{subfigure}{0.25\textwidth}
		\includegraphics[width=\linewidth]{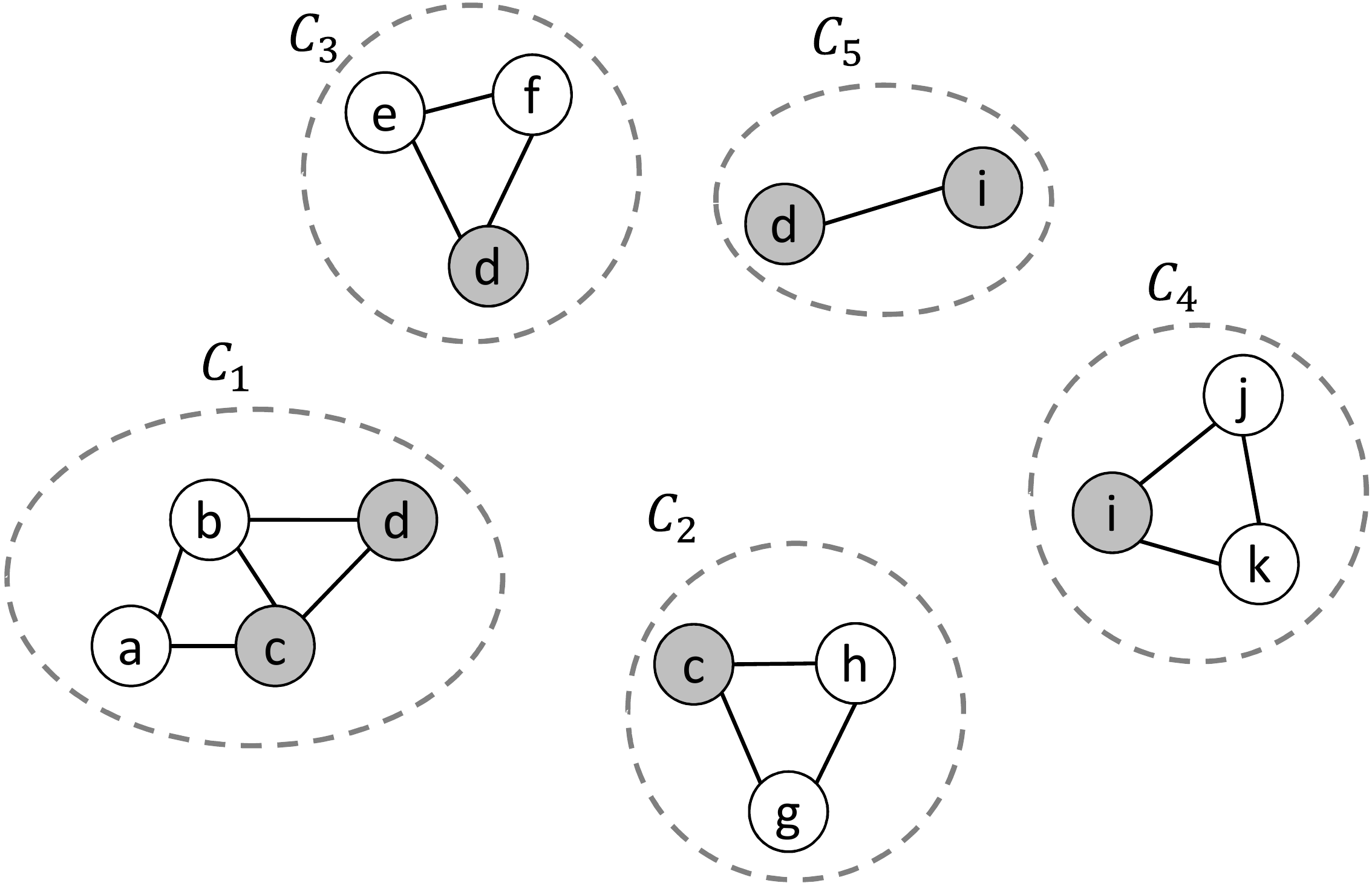}
		\caption{Bi-components.} 
	\end{subfigure}
	\hfill
	\begin{subfigure}{0.25\textwidth}
		\includegraphics[width=\linewidth]{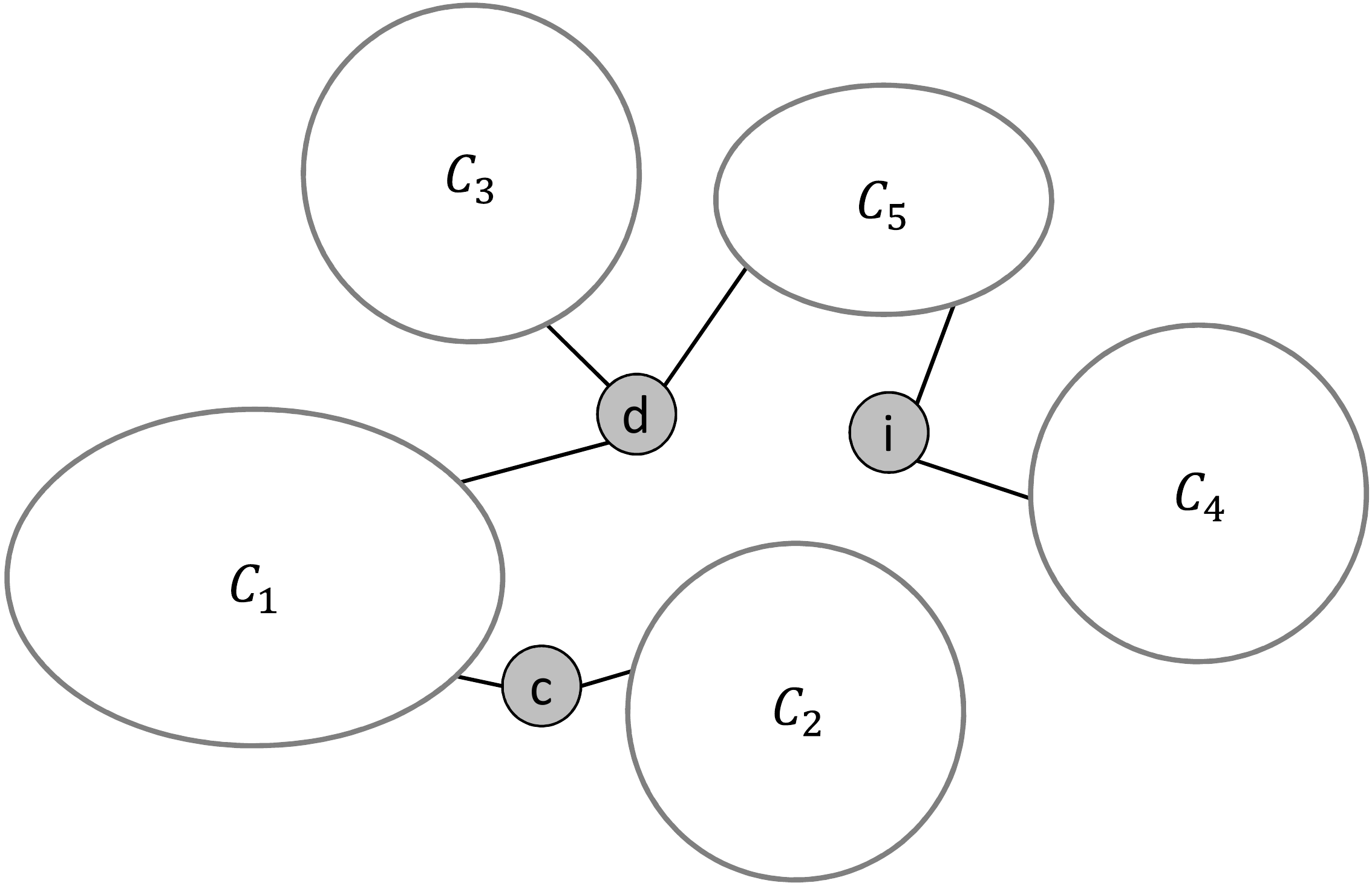}
		\caption{Block-cut tree. } 
	\end{subfigure}	
	\caption{Bi-components in a graph $G$ a) Cutpoints whose removal increase the number of bi-components  b) Five bi-components in $G$ c) Block-cut tree of $G$, created by adding a vertex for each bi-component and each cutpoint in $G$, and adding an edge for each pair of a bi-component and a cutpoint that belongs to that bi-component.}
	\label{fig:net}
	\vspace{-0.4cm}
\end{figure*}
In this section, we describe \RP{} algorithm, an application of \SSP{} framework for RSP$_{bc}$.
We 
start by introducing an auxiliary sample space, call \emph{intra-component shortest path} (\CSP). The \CSP{} sample space is constructed by breaking shortest paths to intra-component shortest paths in which all nodes must belong to the same bi-component. 
In \RP{} algorithm, we select  the exact subspace as the set of all 2-hop shortest paths that go through a node in the subset of target nodes and propose and algorithm that efficiently computes the expected risks on the exact subspace. For the approximate subspace, use the sampling method that is described in Subsection \ref{subsec:estimation} to estimate the expected  risk. We also present a tight bound on the VC dimension that is based on the characteristics of the subset. 


 \vspace{-0.2cm}
\subsection{Sample Space for RSP$_{bc}$}
\label{subsec:bcsamplespace}
 \vspace{-0.1cm}
We start by introducing an auxiliary sample space, termed \emph{intra-component shortest paths} (\CSP), that contains the shortest paths with all nodes belong to the the same bicomponent.  

 Given the target nodes, our sample space will be a ``personalized'' version of the \CSP, obtained by removing shortest paths that have no connections to the target nodes. 
%

\noindent \textbf{\CSP{} sample space.}
The \CSP{} sample space consists of \emph{intra-component shortest paths}, i.e., the shortest path in which source node and the destination node belong to the same \emph{bi-component} \cite{hopcroft1973algorithm}. 
 \vspace{-0.2cm}
\begin{equation*}
\sX_c = \bigcup_{\substack{s',t' \text{ in same bi-component}}}
P(s',t')
 \vspace{-0.2cm}
\end{equation*}

\noindent\emph{Bi-component and block-cut tree. }
A bi-component  is a maximal ``nonseparable'' subgraph, i.e., if any one node is removed, the subgraph still remain connected. 	We denote $\sC = \{C_1,C_2,\cdots,C_\ell\}$ ($\ell \ge 1$) as the set of bi-components of $G$. 
Usually, each node in the graph belongs to exactly one bi-component.
The nodes that belong to more than one bi-component are referred as \emph{cutpoints}. The removal of a cutpoint makes the graph disconnected.
The structure of the bi-components and cutpoints can be described by a tree $G_T=(V_T,E_T)$, called the \emph{block-cut tree} \cite{harary1969matroids}.
The set of nodes $V_T$ consists of all bi-components and cutpoints. Each edge in $E_T$ is a pair of bi-component and a cutpoint that belongs to that bi-component. 
For example, in Fig. \ref{fig:net}, $V_T = \{C_1,C_2,C_3,C_4,C_5,c,d,i\}$ and $E_T = \{(c,C_1),(c,C_2),(d,C_1),(d,C_3),(d,C_5),(i,C_4),(i,C_5)\}$.
 This tree has a node for each bi-component and for each cutpoint of the given graph. There is an edge in the block-cut tree for each pair of a bi-component and a cutpoint that belongs to that component. 
 

\noindent \emph{Breaking a shortest path into intra-component shortest paths.} We construct the \CSP{} sample space by  dividing each shortest path into multiple intra-component shortest paths such that two consecutive intra-component shortest paths belong to different bi-component.  In other words, an intra-component shortest path $p' \in \sX_c$ in a bi-component $C_i$ is a part of a shortest path $p \in \sX_b$ if $p'$ is the maximal intra-component shortest path in $p$ that belongs to the bi-component $C_i$. We denote $I(p)$ as the set of intra-component shortest paths when we break the shortest path $p$. 


The \CSP{} distribution $\dis_c$ over the \CSP{} sample space $\sX_c$ is defined as
\vspace{-0.5cm}
\begin{equation*}
\Pr_{x\sim\dis_c}[x=p'] = \frac{1}{\gamma} \times \Pr_{p\sim\dis_b}[p' \in I(p)],
\end{equation*}
where $\gamma$ is the normalizing factor that will be given in Eq.\ref{eq:normalized} to ensure the sum of all probabilities equals $1$.

\noindent \textbf{\CSP{} distribution.} 
To compute \CSP{} distribution, we introduce the definition of out reach set as follows. 

\noindent \emph{Out reach set.} The out reach set $R_i(v)$ of a node $v$, regarding to a bi-component $C_i$ is the set of all nodes $u$ that can be reach by $v$ without going through any nodes in $C_i$. If $v$ is not a cutpoint, the out reach set $R_i(v)$ only consists of the node $v$ itself. 
If $v$ is a cutpoint, the out reach set $R_i(v)$ consists of the node $v$ itself and all nodes $u$ that belong to a bi-component $C_j$ that can be reach by $v$ in the block-cut tree $G_T$ without going through $C_i$. We can compute the out reach of all cutpoints with the time complexity of $O(|V'|)$ using dynamic programming.

\begin{claim}\label{claim:OUTREACH}
	Consider a bi-component $C_i \in \sC$, each $v \in V$ belongs to exactly one out reach set of a node $u$, regrading to $C_i$

\end{claim}
For Claim \ref{claim:OUTREACH}, for any bi-component $C_i \in \sC$, we have
 \vspace{-0.2cm}
\begin{equation}\label{sums}
\sum_{v \in C_i} r_i(v) = n
 \vspace{-0.2cm}
\end{equation}

Given a bi-component $C_i$, for any pair of node $s' \ne t'$ and $s',t' \in C_i$, let 

\begin{claim}
	\label{claim:st}
	For any intra-component shortest paths $p',p'' \in P_{s't'}$, we have,
	\begin{equation*}
	\Pr_{p\sim\dis_b}[p' \in I(p)] = \Pr_{p\sim\dis_b}[p'' \in I(p)]
	\end{equation*}
\end{claim}

Let 
\vspace{-0.5cm}
\begin{align*}
q_{s't'} 	&= \Pr_{p\sim\dis_b}[p' \in P_{s't'} \text{ and $p' \in I(p)$}]
 \vspace{-0.2cm}
\end{align*}
From Claim \ref{claim:st}, $\forall p' \in P_{s't'}$, we have
 \vspace{-0.2cm}
\begin{align*}
 \Pr_{p\sim\dis_b}[p' \in I(p)]  &= \frac{1}{\sigma_{s't'}} \sum_{p'' \in  P_{s't'}} \Pr_{p\sim\dis_b}[p'' \in I(p)] 
&= \frac{1}{\sigma_{s't'}}q_{s't'} 
 \vspace{-0.2cm}
\end{align*}


\begin{lemma}
	\label{lemma:outreach1}
	Consider an intra-component shortest path $p'$ from $s'$ to $t'$, where $s',t' \in C_i$. If $p'$ is a part of a shortest $p$ from $s$ to $t$, then the $s \in R_i(s')$ and $t \in R_i(t')$ 
\end{lemma}

\begin{lemma}
	\label{lemma:outreach2}
	Consider two node $s',t' \in C_i$. Consider a shortest path $p$ from $s$ to $t$, where $s \in R_i(s')$ and $t \in R_i(t')$. Then, intra-component shortest path $p'$ from $s'$ to $t'$ such that $p'$ is a part of $p$. 
	Consider an intra-component shortest path $p'$ from $s'$ to $t'$. If $p'$ is a part of a shortest $p$ from $s$ to $t$, then the $s \in R_i(s')$ and $t \in R_i(t')$ 
\end{lemma}

From Lemma \ref{lemma:outreach1} and  Lemma \ref{lemma:outreach2}, we have
 \vspace{-0.2cm}
 $$q_{s't'} = \frac{1}{n(n-1)} \mid R_i(s')\mid \times \mid R_i(t')\mid = \frac{1}{n(n-1)} r_i(s') r_i(t'). \vspace{-0.2cm}$$

Let $\gamma$ be the sum of $q_{st}$ for all pair $s,t$, i.e.,
 \vspace{-0.2cm}
\begin{align}
\label{eq:normalized}
	\gamma = \sum_{i=1}^\ell\sum_{s \ne t\in C_i} q_{st}	= \frac{1}{n(n-1)}\sum_{i=1}^\ell\sum_{s \in C_i} r_i(s) (n - r_i(s)) 
	 \vspace{-0.2cm}
\end{align}
To compute $\gamma$, we iterate through all bi-components and for each bi-component $C_i$, we iterate through all nodes in $C_i$. This can be done in $O(n)$.

For an intra-component shortest path $p'$ from $s'$ to $t'$, we have
 \vspace{-0.2cm}
\begin{equation}
\Pr_{x\sim \dis_c}[x=p'] = \frac{1}{\gamma} \frac{1}{\sigma_{s't'}}q_{s't'}
\label{eq:mu}
 \vspace{-0.2cm}
\end{equation}


\noindent \emph{Betweenness centrality for cutpoints.} 
When we break a shortest path into multiple intra-component shortest paths, the target node of the previous intra-component shortest path is the source node of the next intra-component shortest path. We refer to those nodes as \emph{break points}. Since the break points belong to more than one bi-component, they must be cutpoints. 

The break points were inner nodes in the original shortest path but they are not accounted when we compute the betweenness centrality on the \CSP{} sample space. 

For a cutpoint $v$, let $bc(v)$ be the probability that $v$ is a break point of a shortest path $p \in \sX_b$. 
 \vspace{-0.2cm}
\begin{align*}
bc_a(v) &= \Pr_{p \sim \dis_b}[v \text{ is a break point of } p].
 \vspace{-0.2cm}
\end{align*}

\begin{lemma}
	\label{lemma:bc}
	For any node $v \in V$, 
	 \vspace{-0.2cm}
	\begin{equation*}
	bc(v) =\mathds{E}_{p \sim\dis_b}g(v,p) = {\gamma}\mathds{E}_{p \sim\dis_c}g(v,p) + 	bc_a(v),
	 \vspace{-0.2cm}
	\end{equation*}
	
	where $g(v,p) = 1$ if $v$ is an inner node of $p$ (see Eq. \ref{eq:gv}) 
\end{lemma}



Next, we show how to compute the value of $bc_a(v)$ for a cutpoint $v$. 
Consider the block-cut tree $G_T$ and take $v$ as the root node of $G_T$.
By removing the root node $v$, we can divide $G_T$ into multiple subtrees where the root nodes of those subtrees are the bi-components that $v$ belongs to. The node $v$ is a break point of a shortest path $p$ from $s$ to $t$ if the source node $s$ and the target node $t$ belong to two difference subtrees.

Formally, let $T_i(v)$ be the set of nodes (except $v$) that belong a bi-component in the subtree in which $C_i$ is the root node. 
We have,
 \vspace{-0.2cm}
\begin{align*}
T_i(v) &= V \setminus R_i(v) \text{ and }
\bigcup_{i \in [\ell]: v \in C_i} T_i(v) &= V \setminus \{v\}.
 \vspace{-0.2cm}
\end{align*}

\begin{lemma}
	\label{lemma:bca}
	A node $v \in V$ is a break point of a shortest path $p$ from $s  \in T_i(v)$ to $t  \in T_i(j)$
	if  $i \ne j$. 
\end{lemma}

%

From Lemma \ref{lemma:bca}, we have, 
 \vspace{-0.2cm}
\begin{align}
\nonumber bc_a(v)
&= \frac{1}{n(n-1)}\sum_{C_i \in \sC| v \in C_i} |T_i(v)| \sum_{C_j \in \sC\setminus\{C_i\}| v \in C_j} |T_j(v)| \\
&= \frac{1}{n(n-1)}(r_{i}(v) - 1)  (n - r_{i}(v))
\label{eq:bca}
 \vspace{-0.2cm}
\end{align}




%

\noindent \textbf{Personalized ISP (\PCSP{}) Sample Space.}    Given a subset $A \subseteq V$, 
we construct a personalized sample space by taking the necessary 
samples in the \CSP{} sample space.
 Specifically, 
the personalized \CSP{}  sample space $\sX_c^{(A)}$ consists of all shortest paths from $s$ to $t$ such that both $s, t$
belong to 
some bi-component $C_j$ that contains at least one node in $A$. 



Formally, let $I(A) = \{i \in [\ell]:  C_i\cap A \ne \emptyset\}$
be the set of the index of bi-components that contains at least one node in the subset $A$. We have 
 \vspace{-0.2cm}
\begin{equation}
\label{eq:pisp}
\sX_c^{(A)} = \bigcup_{i \in I(A)}\bigcup_{\substack{s' \ne t' \in C_i}}
P(s',t')
 \vspace{-0.2cm}
\end{equation} 

Let $\eta$ be the probability that a shortest path in the \CSP{} sample space $\sX_c$ belongs to the  \PCSP{} sample space $\sX_c^{(A)}$, i.e., 
 \vspace{-0.2cm}
\begin{align}
\label{eq:pnormalized}
\eta &= \Pr_{x\sim\dis_c}[x \in \sX_c^{(A)}]
= \frac{\sum_{i \in I(A)}\sum_{s \ne t \in C_i}q_{st}}{\sum_{i=1}^\ell\sum_{s \ne t\in C_i} q_{st}}\\
\nonumber&= \frac{\sum_{i \in I(A)}\sum_{s \in C_i} r_i(s) (n - r_i(s))}{\sum_{i=1}^\ell\sum_{s \in C_i} r_i(s) (n - r_i(s))}
 \vspace{-0.2cm}
\end{align}
We can compute $\eta$ in $O(n)$ (similar with computing $\gamma$).

We define The \PCSP{} distribution $\dis_c^{(A)}$ over the \PCSP{} sample space $\sX_c^{(A)}$ as follows. For any  $i \in I(A), \forall (s,t) \in C_i, \forall p \in P_{st}$,
 \vspace{-0.4cm}
\begin{equation}
\label{eq:pispdis}
\Pr_{x\sim\dis_c^{(A)}}[x=p'] = \frac{1}{\eta} \times \Pr_{x\sim\dis_c}[x=p'] = \frac{1}{\sigma_{st}}\frac{q_{st}}{\gamma\eta},
 \vspace{-0.2cm}
\end{equation}

\begin{lemma}
	\label{lemma:bcdis}
	For any node $v \in A$, we have,
	 \vspace{-0.2cm}
	\begin{equation}
		bc(v) = {\gamma} \eta \mathds{E}_{p \sim\dis_c^{(A)}}g(v,p) + bc_a(v)
		\label{eq:bcdis}
		 \vspace{-0.2cm}
	\end{equation}
\end{lemma}

\subsection{Sample space partitioning for RSP$_{bc}$}


Now, we show how to model the betweenness centrality ranking problem as a hypothesis ranking problem and how to apply the \SSP{} framework.

Here, we consider a binary  label space 
 \vspace{-0.2cm}
\begin{equation}
\label{eq:labelspace}
\sY_c = \{0,1\},
 \vspace{-0.2cm}
\end{equation}
 the labeling function $\lf_c$ always returns $0$, i.e., 
  \vspace{-0.2cm}
\begin{equation}
\label{eq:lf}
\lf_c(\sx) = 0, \forall \sx \in \sX_c^{(A)},
 \vspace{-0.2cm}
\end{equation}
and the hypothesis class
 \vspace{-0.2cm}
\begin{equation}
\label{eq:hypotheses}
	\sH_c^{(A)} = \{h_v \defeq g(v,\cdot)\}_{v \in A}
	 \vspace{-0.2cm}
\end{equation}
where $g$ is given in Eq. \ref{eq:gv}, i.e., $h_v(p) = 1$ if $v$ is an inner node of $p$. 

As $\lf_c = 0, \forall \sx \in \sX_c^{(A)}$, we have, 
$\loss(h_v(x),\lf_c(x)) = h_v(\sx).$
We denote $\risk_c^{(A)}(\sh_v)$ as the expected risk of the hypothesis $h_v$, i.e.,
 \vspace{-0.2cm}
\begin{align*}
\risk_c^{(A)}(\sh_v) &= \sum_{p \in \sX_c^{(A)}} \Pr_{x\sim \dis_c^{(A)}}[x=p]\loss\left(\sh_v(p),\lf_c(p)\right)
 \vspace{-0.2cm}
\end{align*}

 \vspace{-0.2cm}
\begin{lemma}
	For any node $v \in A$, we have,
	\begin{equation*}
		bc(v) = {\gamma} \eta \risk_c^{(A)}(\sh_v) + bc_a(v)	
	\end{equation*}
	\label{lemma:bcc}
\end{lemma}
 \vspace{-0.5cm}



Following \SSP{} framework, we divide $\sX_c^{(A)}$ into exact subspace and approximate subspace.\\ 
\noindent \textbf{Exact subspace.}
\label{subsec:bcexact}
We choose the exact subspace is the set of all shortest paths $p$ that have the length equals $2$ ($len(p) = 2$) and there exists a node $v \in A$ such that $v$ is an inner node of $p$ ($g(v,p) = 1$).
 \vspace{-0.2cm}
\begin{equation}
\hat{\sX}_c^{(A)} = \{ p \in \sX_c^{(A)} |\ len(p) = 2 \text{ and } \exists v \in A \text{ s.t. } g(v,p) = 1 \}
\label{eq:exact-subspace}
\end{equation}

For a node $v \in A$, the expected risk of $h_v$ on the exact subspace is computed as follows
 \vspace{-0.1cm}
\begin{equation*}
\hat{\erisk}_v = \sum_{p \in \hat{\sX}_c^{(A)}} \Pr_{x\sim \dis_c^{(A)}}[x=p]\loss\left(\sh_v(p),\lf_c(p)\right)
 \vspace{-0.2cm}
\end{equation*}

Here, we present the \exactbc{} algorithm to efficiently compute the expected risks on the exact subspace.
Let $B$ be the set of all neighbors of nodes in $A$. 
For each bi-component $C_i$, for each source node $s \in B \cap C_i$, we execute two phases as follows. In the first phase,
 we find all
the shortest paths of length $2$ from $s$ to $t \in  B \cap A$. 
Let $\Delta_s$ be the set of nodes $t$ such that the distance from $s$ to $t$ is $2$ (i.e., $d_{st} = 2$). 
For a node $t \in \Delta_s$, we denote $w_t$ as the number of shortest paths from $s$ to $t$. Initially, all we set $w_t = 0$, for all $t \in B$. To find the value of $w_t$, we iterate through all neighbors $v$ of $s$, then iterate through all neighbors $t$ of $v$. If $t$ is not a neighbor of $s$, i.e., $d_{st} = 2$, we add $t$ to $\Delta_s$ and increase the value of $w_t$ by $1$.
In the second phase,
 we calculate the two-hop expected risks on the exact subspace of all nodes $v \in A$ based on the number of shortest paths that we found in the first phase. 
 Due to the space limitation, we present the pseudocode of \RP{} algorithm in the Appendix of the full version \cite{fullversion}.
 \vspace{-0.2cm}
\begin{lemma}
	\label{LEMMA:2HOP}
	Let $\{\hat{\erisk}_v\}_{v \in A}$ and $\hat{Q}/Q$ be the output of 
	Algorithm \exactbc{}. For all node $v \in A$, we have
	 \vspace{-0.2cm}
	\begin{align*}
&\hat{\erisk}_v = \sum_{\sx \in \hat{\sX}} \Pr_{x\sim \dis_c^{(A)}}[x=p]\loss\left(\sh_v(p),\lf_c(p)\right)\\
&\Pr_{\sx\sim\dis_c^{(A)}}[\sx \in \hat{\sX}_c^{(A)}] = \hat{\lambda}
 \vspace{-0.2cm}
	\end{align*}
\end{lemma}

\begin{lemma}
	\label{LEMMA:COMPLEXITY-subset}
	Algorithm \exactbc{} has the  time complexity of $O(K)$, where 
	 \vspace{-0.2cm}
	$$K = \sum_{v \in B} deg(v)^2. \vspace{-0.2cm}$$
	Here, $deg(v)= |Adj(v)|$ is the degree of $v$.
\end{lemma}

Note that, 
the expected risks on the exact subspace provide ``non-empty'' estimations for the expected risks. 
 \vspace{-0.2cm}
\begin{lemma}
	For all node $v \in V$, we have
	 \vspace{-0.2cm}
	\begin{equation*}
	\text{If } \risk_c^{(A)}(\sh_v) > 0, \text{then } \hat{\erisk}_v > 0. 
	 \vspace{-0.2cm}
	\end{equation*}
	\label{lemma:falsezero}
\end{lemma}
 \vspace{-0.2cm}
%
%
 \vspace{-0.2cm}
\noindent \textbf{Approximate subspace.}
\label{subsec:bcestimate}
The approximate subspace  is  the set of the remaining shortest paths after removing the shortest path in the exact subspace, i.e.,
 \vspace{-0.2cm}
\begin{equation}
\tilde{\sX}_c^{(A)} = \sX_c^{(A)} \setminus \hat{\sX}_c^{(A)} 
\label{eq:estimation-subspace}
 \vspace{-0.2cm}
\end{equation}
We define the distribution $\tilde{\dis}_c^{(A)}$ over the approximate subspace $\tilde{\sX}_c^{(A)} $, where the probability to select a path $p'$ from $s'$ to $t'$ is 
 \vspace{-0.2cm}
\begin{align}
\label{eq:tmu}
\Pr_{x\sim \tilde{\dis}_c^{(A)}}[x=p'] &=  \frac{1}{1-\hat{\lambda}} \Pr_{x\sim {\dis}_c^{(A)}}[x=p'] 
= \frac{1}{1-\hat{\lambda}} \frac{1}{\gamma} \frac{1}{\sigma_{s't'}}q_{s't'}
 \vspace{-0.2cm}
\end{align}

The expected risk on the approximate subspace $\tilde{\sX}_c^{(A)}$ is computed as follows
		\vspace{-0.2cm}
\begin{equation}
\riskest_c^{(A)}(h_v) = 
\sum_{x \in \tilde{\sX}_c^{(A)}} \Pr_{x\sim \tilde{\dis}_c^{(A)}}[x=p]\loss\left(\sh_v(x),\lf_c(x)\right)
		\vspace{-0.2cm}
\end{equation}


\subsection{Risk Estimation in the Approximate Space}
We use the same techniques in Subsection \ref{subsec:estimation} to estimate the expected risk of the hypotheses in the approximate space. 
Here, we present an algorithm to generate samples in the approximate space and show a bound VC dimension, thus, obtaining a stronger bound on sample complexity. 

\noindent \textbf{Generating samples.} We use rejection sampling and  multistage sampling techniques to generate samples in the approximate space. 

\noindent
\emph{Rejection sampling.} We apply a rejection sampling method to sample a shortest path $p$ from $\tilde{\sX}_c^{(A)}$ by repeating sampling a shortest path $p$ from $\sX_c^{(A)}$ until $p \notin \hat{\sX}_c^{(A)}$.
\begin{algorithm}[ht!]
	
	\SetKwInOut{Input}{Input}
	\SetKwInOut{Output}{Output}	
	\Input{A graph $G$, its set of bi-components $\sC$, and a subset $A$}
	\Output{A sample $p$ as a random shortest path in the estimation subspace $\tilde{\sX}_c^{(A)}$ with the probability propositional to the probability distribution $\sD_c^{(A)}$}
	\Repeat{$p \notin \hat{\sX}_c^{(A)}$}{
		Pick a $i \in I(A)$ randomly with probability $\Pr_{x \sim \dis_c^{(A)}}[x \in C_i]= \frac{n^2 - \sum_{s \in C_i}r_i(s)^ 2}{\gamma\eta}$ \\ 
		Pick a source node $s \in C_i$ randomly with probability $\Pr[s| x \in C_i]  = \frac{r_i(s)(n-r_i(s))}{n^2 - \sum_{s \in C}r_i(s)^ 2}$ \\ 
		Pick a target node $t\in C_i \setminus \{s\}$ randomly with probability $\Pr[t|s] =  \frac{ r_i(t)}{n-r_i(s)}$ \\ 
		Uniformly pick a shortest path $p$ from $s$ to $t$, i.e., $\Pr[p|st] = \frac{1}{\sigma_{st}}$;
	}        
	Return $p$
	\caption{Algorithm \samplebc}
	\label{alg::dbpath-sampling}	
\end{algorithm} 

\noindent
\emph{Multistage sampling.} We use a multistage sampling method to reduce the space complexity of $O(n)$. Our multistage sampling method consists of $4$ steps (please see Algorithm \ref{alg::dbpath-sampling}). First, we pick a bi-component $C_i$ with probability  $\Pr_{x \sim \dis_c^{(A)}}[x \in C_i]$. Secondly, we pick a source node $s \in C_i$ with probability $\Pr[s| x \in C_i]$. Thirdly, we pick a target node $t \in C_i \setminus\{s\}$ with probability $\Pr[t|s]$. Finally, we pick a shortest path $p$ from $s$ to $t$ with probability $\Pr[p|st]$. 
By using the multistage sampling method as above, the probability that we pick a shortest path $p$ from $s$ to $t$ in a bi-component $C_i$ is
 \vspace{-0.2cm}
\begin{equation*}
\Pr_{x \sim \dis_c^{(A)}}[x \in C_i] \times \Pr[s| x \in C_i] \times \Pr[t|s] \times \Pr[p|st] = \frac{1}{\sigma_{st}}\frac{q_{st}}{\gamma\eta}
 \vspace{-0.2cm}
\end{equation*}

To uniformly sample a shortest path $p$ from $s$ to $t$, we perform a balanced bidirectional BFS (breadth-first search) \cite{Borassi16} to find all shortest paths from $s$. We execute two BFSs from both the source node $s$ and the target node $t$, in such a way that the two BFSs are likely to explore about the same number of edges. When the two BFSs ``touch each other'', we can obtain the distance and all the shortest paths from $s$ to $t$.

 \vspace{-0.2cm}
\begin{lemma}
	\label{LEMMA:SAMPLE}
	The probability that algorithm \samplebc{} 
	returns a shortest path $p'$  from $s'$ to $t'$ with probability
 \vspace{-0.2cm}
	$$\Pr_{x\sim \tilde{\dis}_c^{(A)}}[x=p'] = \frac{1}{1-\hat{\lambda}} \frac{1}{\gamma} \frac{1}{\sigma_{s't'}}q_{s't'}.  \vspace{-0.2cm}$$
\end{lemma}

Borassi et al. \cite{Borassi16} analyze the time complexity of balanced bidirectional BFS in a random graph as follows.
%
 \vspace{-0.2cm}
\begin{lemma} [Theorem 4 \cite{Borassi16}]
	\label{lemma:samplecomp}
	Let $G$ be a graph generated through the aforementioned models\cite{Borassi16}. 
	For each pair of nodes $s,t$, w.h.p., the time needed to compute an $st$-shortest path through a bidirectional BFS is $\O(n^{\frac{1}{2}+o(1)})$ if the degree distribution has finite second moment. 
\end{lemma}

\begin{table*}[htp!]
	\centering
	\begin{tabular}{ c|c|c|c }
		Subset & Full network & Any subset $A$ & $l$-hop neighbors\\
		\hline
		Riondato et al.\cite{Riondato16}  &$\lfloor \log(\VD(V)-1) \rfloor + 1$& $\lfloor \log(\VD(V)-1) \rfloor + 1$ & $\lfloor \log(\VD(V)-1) \rfloor + 1$\\
		\RP &$\lfloor \log(BD(V)-1) \rfloor + 1$ & $\lfloor \log(BS(A)) \rfloor + 1$ &  $\lfloor \log(2l+1) \rfloor + 1$
	\end{tabular}
\vspace{-0.1cm}
	\caption{The comparison on the bound of the VC-dimension. Here $\VD(V)$ is the diameter of the graph, $BD(V)$ is the maximum diameter of a bicomponent of the graph, and $BS(A)$ is the maximum number of nodes $A$ that appear in the same shortest path. 
		\label{table:vc}}
	\vspace{-0.5cm}
\end{table*}
\noindent \textbf{Personalized VC dimension and Sample Complexity.}
Here, we show the analysis for the VC dimension on the personalized \CSP{} sample space. 
By using the bi-component-based sampling method, we can reduce the VC-dimension from log of the diameter of the graph \cite{Riondato16} to log of maximum diameter of a bi-component in the graph. Further, for a specific subset of nodes, we can further reduce the VC-dimension based on the properties of the subset.

For a shortest path $p \in \sX_c^{(A)}$, let $\pi(p)$ be the number of hypotheses $h_v \in \sH_c^{(A)}$ such that $h_v(p) = 1$. From Eq. \ref{eq:gv}, $h_v(p) = 1$ iff $v$ is an inner node in $p$. Thus, $\pi(p)$  is the number of nodes in $A$ that are inner nodes of $p$. 
Recall that, in Lemma \ref{lemma:vc}, we have shown that $VC(\sH_c^{(A)}) \leq \lfloor \log(\pi_{\max}) \rfloor + 1$, where $\pi_{\max} = \max_{x\in\sX} \pi(x)$. 
Thus, we have the following corollary. 
 \vspace{-0.2cm}
\begin{corol}
	Let $BS(A)$ be the maximum number of nodes in $A$ that are inner nodes of a shortest path in $\tilde{\sX}_c^{(A)}$. Let $\sH_c^{(A)}$ is defined as in Eq. \ref{eq:hypotheses}
	We have, 
	\vspace{-0.2cm}
	\begin{equation}
	VC(\sH_c^{(A)}) \leq \lfloor \log(BS(A)) \rfloor + 1
	\label{eq:vc}
	 \vspace{-0.2cm}
	\end{equation}
\end{corol}

%

Note that, it is expensive to compute the exact value of $BS(A)$. Thus, we bound the value of $BS(A)$ as follows. 

\begin{lemma}
	For a subset of nodes $A' \subseteq V$, let $\VD(A') = \max_{s,t \in A'}d_{st}$ be the diameter of $A'$. We have, 
	 \vspace{-0.2cm}
	\begin{align}
	BS(A) \le \max_{i=1}^\ell \left(\min(\VD(C_i)-1,\VD(A\cap C_i),  +1,|A \cap C_i|)\right)
	\label{eq:BS}
	 \vspace{-0.3cm}
	\end{align}
	\label{lemma:BS}
\end{lemma}
 \vspace{-0.3cm}
We can simplify the bound for $BS(A)$ 
based on the maximum diameter of a bi-component 
 \vspace{-0.2cm}
\begin{equation}
	BD(V) = \max_{i=1}^\ell \VD(C_i)
	 \vspace{-0.2cm}
\end{equation}
and the maximum distance between any two nodes in the same bi-component in $A$
 \vspace{-0.3cm}
\begin{equation}
SD(A) =  \max_{i=1}^\ell \VD(A\cap C_i).
 \vspace{-0.3cm}
\end{equation}

Indeed, from Eq. \ref{eq:BS}, we have, 
 \vspace{-0.3cm}
\begin{align*}
	BS(A) 
	&\le \min (  \max_{i=1}^\ell \VD(C_i) -1, \max_{i=1}^\ell \VD(A\cap C_i) + 1) \\
	&= \min (BD(V) - 1,SD(A) + 1)
	 \vspace{-0.3cm}
\end{align*}

%

We bound the diameter of a subset of nodes $A'$ as follows. We pick a random source nodes $s$ and perform a breadth-first search \cite{cormen2009introduction} to find the distance from $s$ to all other node in $A'$. The diameter of $A'$ cannot be bigger than double of the maximum distance from $s$ to a node $t \in A'$, i.e.,
 \vspace{-0.2cm}
\begin{align*}
\forall s \in A', \VD(A') \le 2 \max_{t \in A'}d_{st}
 \vspace{-0.2cm}
\end{align*}

\vspace{-0.2cm}
\noindent {\emph{Comparison on the bound of the VC-dimension.}}
In Table \ref{table:vc}, we compare the VC-dimension of \RP{} and the work in \cite{Riondato16}. In Riondato et al.\cite{Riondato16}, the VC-dimension equals on log of 
the diameter $VD(V)$ of the network. In \RP{}, by using the bi-component-based sampling method, on the full network, the VC-dimension reduces to log of  the maximum diameter $BD(V)$ of a bi-component in the network. On a subset $A$, the VC-dimension further reduces to log the maximum distance between two nodes in $A$ that belong to the same bi-component. Specifically, if $A$ is a subset of $l$-hop neighbors of a node $v$, the VC-dimension equals log of $2l+1$.

 \vspace{-0.2cm}
\subsection{\RP{} algorithm}
\label{subsec:rp}
 \vspace{-0.1cm}
We now describe \RP{} algorithm. 
 At the beginning, we decompose graph $G$ into bi-components $\{C_1,\ldots, C_\ell\}$ and compute that out reach for each node. This can be done in $O(m+n)$. 
We define $\sX_c^{(A)},  \lf_c, \dis_c^{(A)}, H_c^{(A)}$  as in Eq. \ref{eq:pisp}, Eq. \ref{eq:lf}, Eq. \ref{eq:pispdis}, Eq. \ref{eq:hypotheses}, respectively. The sample space $\sX_c^{(A)}$ is partitioned into $\hat{\sX}_c^{(A)} \cup \tilde{\sX}_c^{(A)}$ where 
\begin{align*}
\hat{\sX}_c^{(A)} &= \{ p \in \sX_c^{(A)} |\ len(p) = 2 \text{ and } \exists v \in A \text{ s.t. } g(v,p) = 1 \} \\
\tilde{\sX}_c^{(A)} &= \sX_c^{(A)} \setminus \hat{\sX}_c^{(A)}
\end{align*}
Then, we compute  $\gamma, \eta$ as in Eq. \ref{eq:normalized}, Eq. \ref{eq:pnormalized}, respectively. The computation of $\gamma, \eta$ can be done in $O(n)$. 
Let $\epsilon^* = \epsilon \gamma \eta$. We obtain the estimation $\{\erisk_v\}_{v\in A}$ by running \SSP{} with input $(\sX_c^{(A)},\sD_c^{(A)},\lf_c,\sH_c^{(A)},\epsilon^*,\delta)$, a partition $\sX_c^{(A)} = \hat{\sX}_c^{(A)} \cup \tilde{\sX}_c^{(A)}$. In \RP algorithm, we use algorithm \exactbc{} to compute the compute the expected risks on the exact subspace, and algorithm \samplebc{} to generate a sample.

For each node $v \in V$, we compute $ bc_a\left(v\right)$ as in Eq.\ref{eq:bca} as output an estimation for the betweenness centrality \vspace{-0.2cm}
$$\tilde{bc}(v) = bc_a\left(v\right) + \gamma \eta \erisk_v.\vspace{-0.2cm}$$

Due to space limitations, we present the pseudocode of \RP{} algorithm in the Appendix of the full version \cite{fullversion}.
\subsection{Correctness and Complexity.}

\begin{theorem}
	\label{THEOREM:BC}
	Let $\{\tilde{bc}(v)\}_{v \in A}$ be the estimation that is returned by \RP{} algorithm. We have
\vspace{-0.2cm}
		$$\Pr \left[\forall v \in A, \tilde{bc}(v) - bc(v)| < \epsilon\right] \ge 1-\delta. \vspace{-0.2cm}$$
\end{theorem}


\begin{lemma}
	\label{lemma:rt}
		Let $G$ be a graph generated through the aforementioned models\cite{Borassi16}. 
		\RP{} algorithm has a time complexity of $O(m+n + K + \frac{1}{\epsilon^2}(\lfloor \log(BS(A)) \rfloor + 1+ln\frac{1}{\delta}) n^{1/2 + o(1)})$, 
\end{lemma}

Note that, due to the space limitation, here, we omit the proofs of the lemmas. The detailed proofs are presented in the Appendix of the full version \cite{fullversion}.

%% file: Sections/6_Experiments.tex
 \vspace{-0.2cm}
{
\section{Experiments}
}
 \vspace{-0.2cm}

\begin{figure*}[!ht]
	\centering 
	\begin{subfigure}{0.5\textwidth}
		\includegraphics[width=\linewidth]{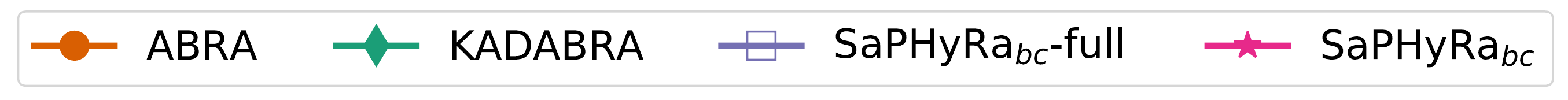}
	\end{subfigure}
	
	\begin{subfigure}{0.18\textwidth}
		\includegraphics[width=\linewidth]{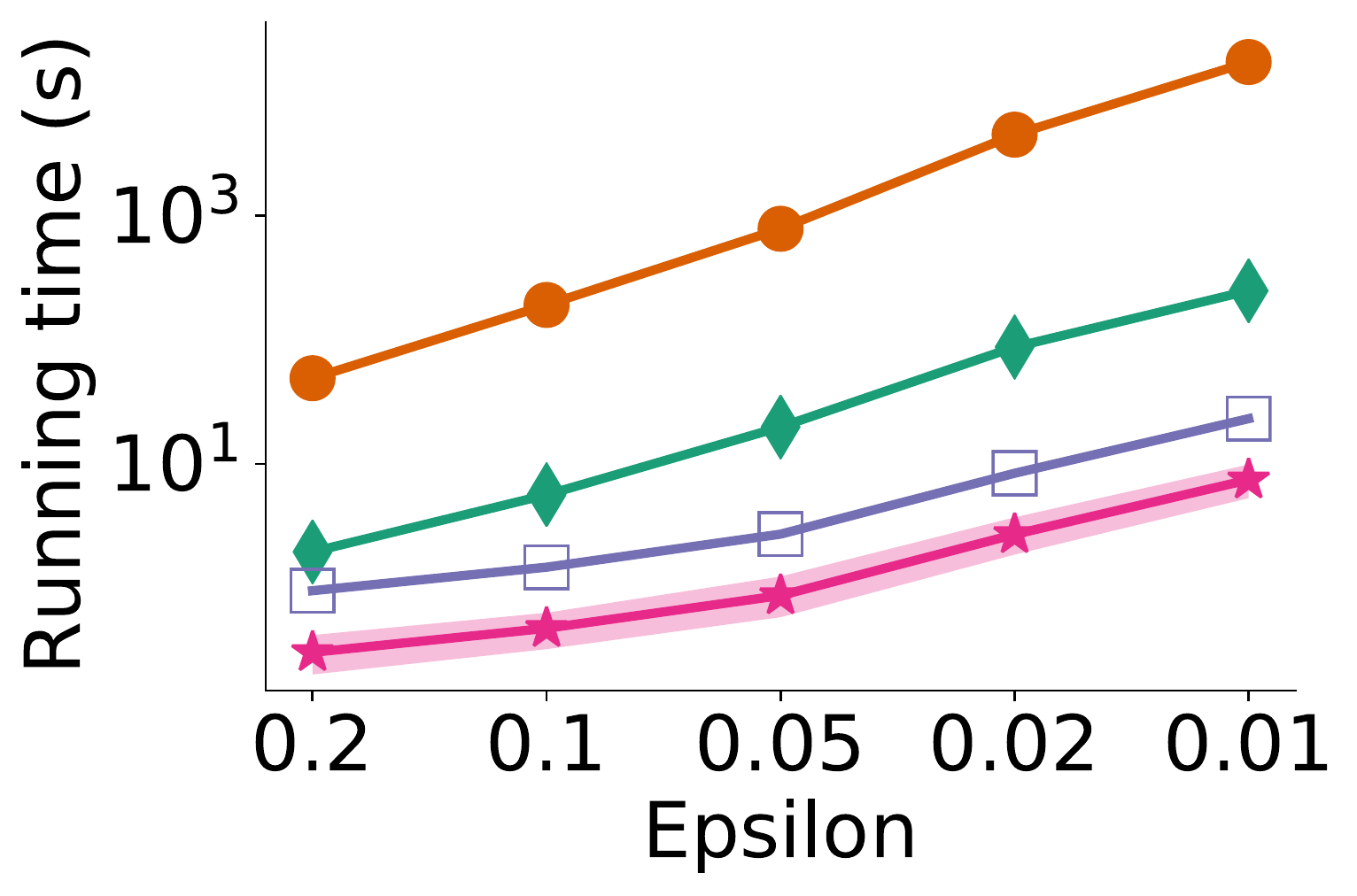}
		\vspace{-0.2cm}
		\caption{Flickr} 
	\end{subfigure}	
	\begin{subfigure}{0.18\textwidth}
		\includegraphics[width=\linewidth]{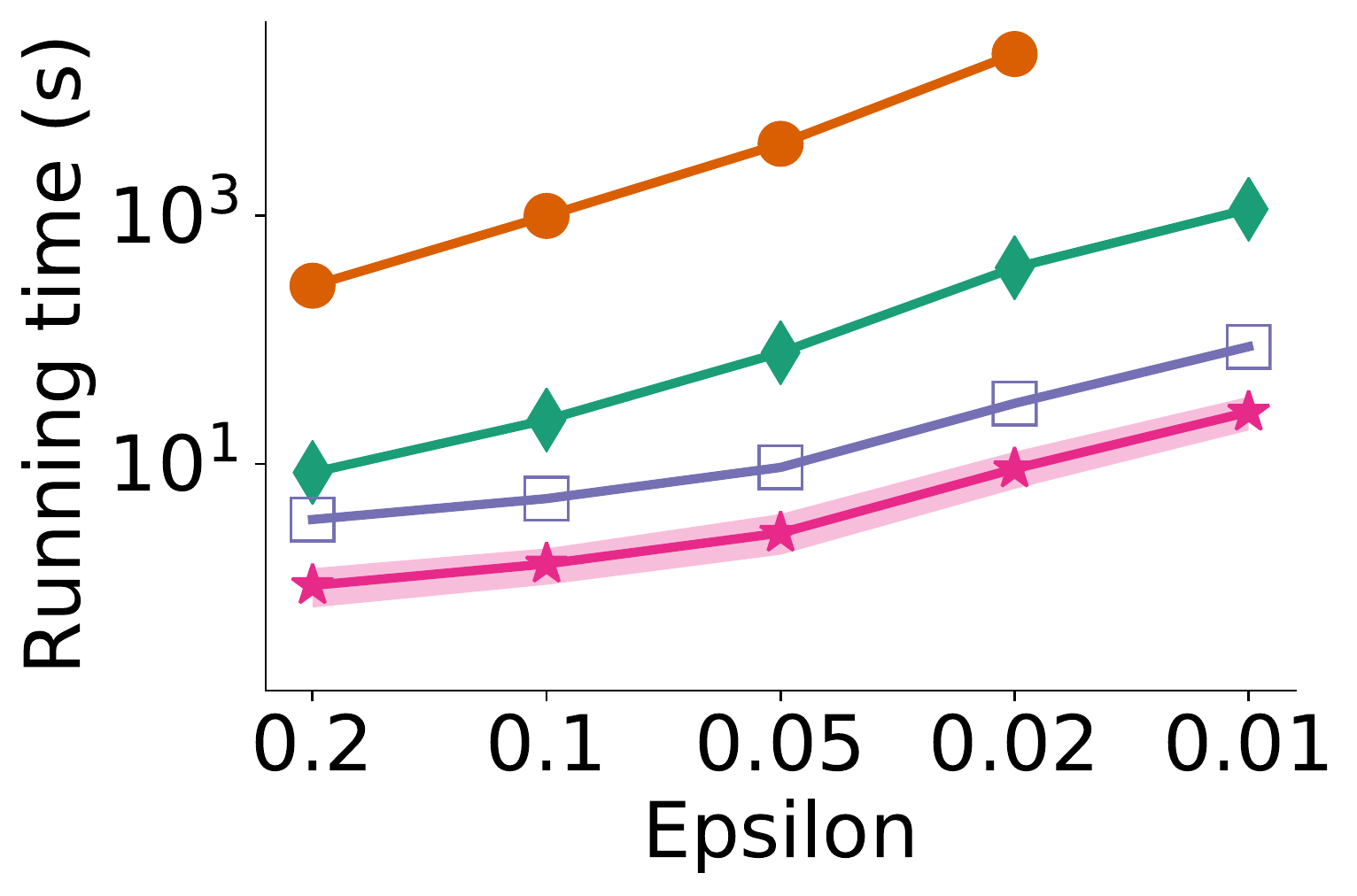}
		\vspace{-0.2cm}
		\caption{LiveJournal} 
	\end{subfigure}	
	\begin{subfigure}{0.18\textwidth}
		\includegraphics[width=\linewidth]{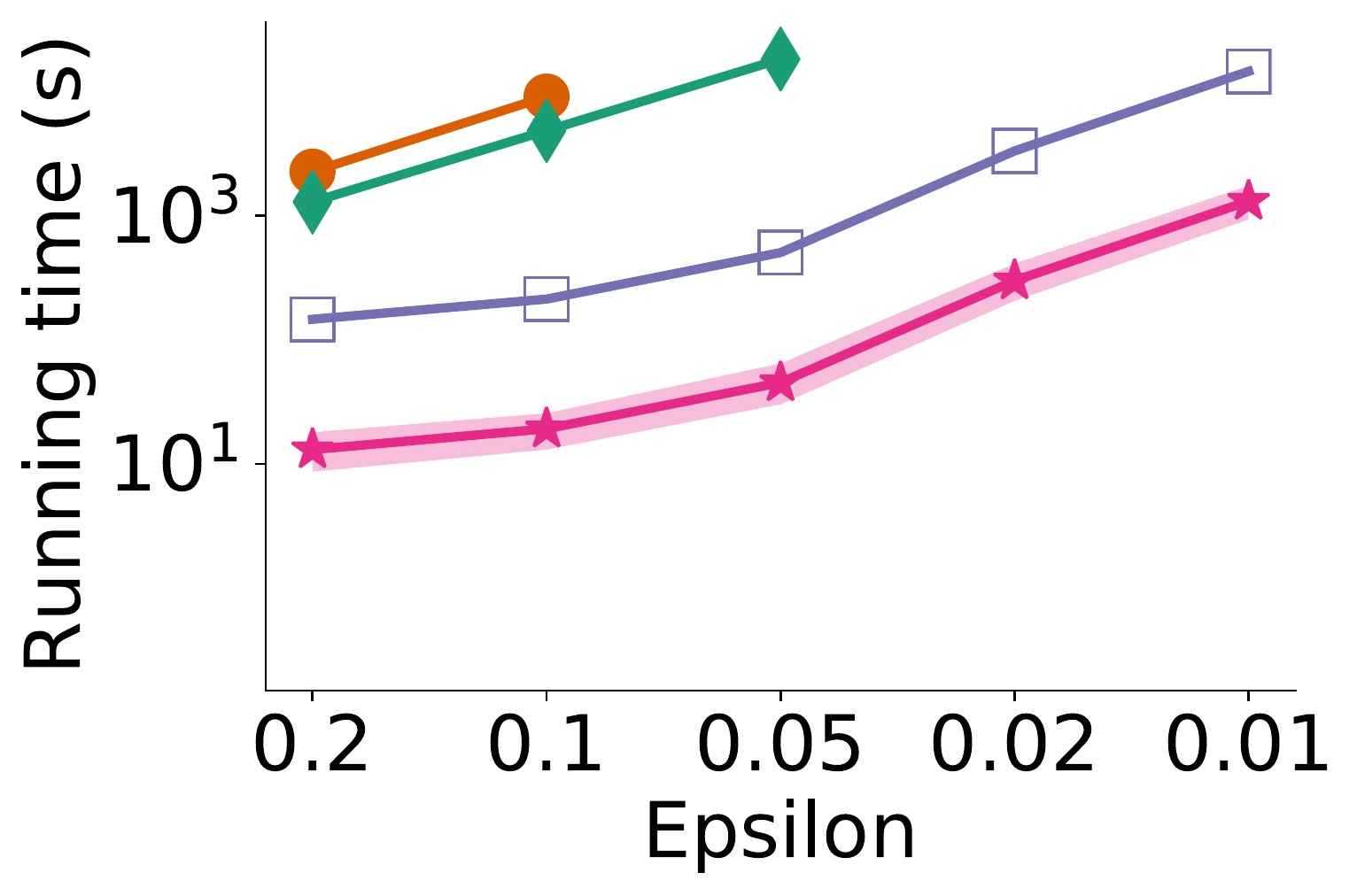}
		\vspace{-0.2cm}
		\caption{USA-road} 
	\end{subfigure}	
	\begin{subfigure}{0.18\textwidth}
		\vspace{-0.2cm}
		\includegraphics[width=\linewidth]{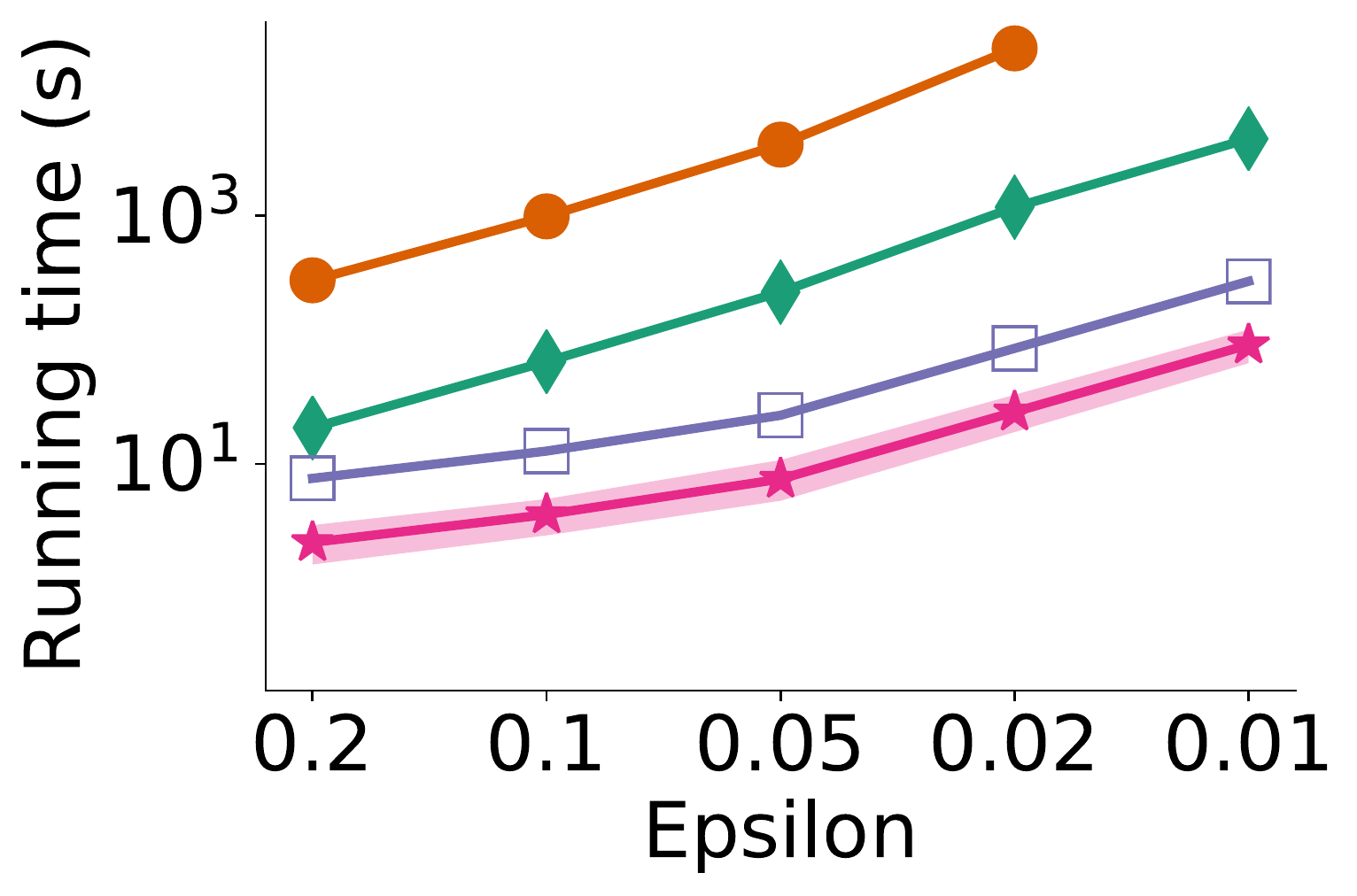}
		\caption{Orkut} 
	\end{subfigure}	
	\vspace{-0.2cm}
	\caption{\small Running time (log-scale). The shaded areas show the $95\%$ confident intervals of  \SSP{} over different target subsets. }
	\label{fig:runningtime}
	\vspace{-0.6cm}
\end{figure*} 

\begin{figure*}[!ht]
	\centering 
	\begin{subfigure}{0.18\textwidth}
		\includegraphics[width=\linewidth]{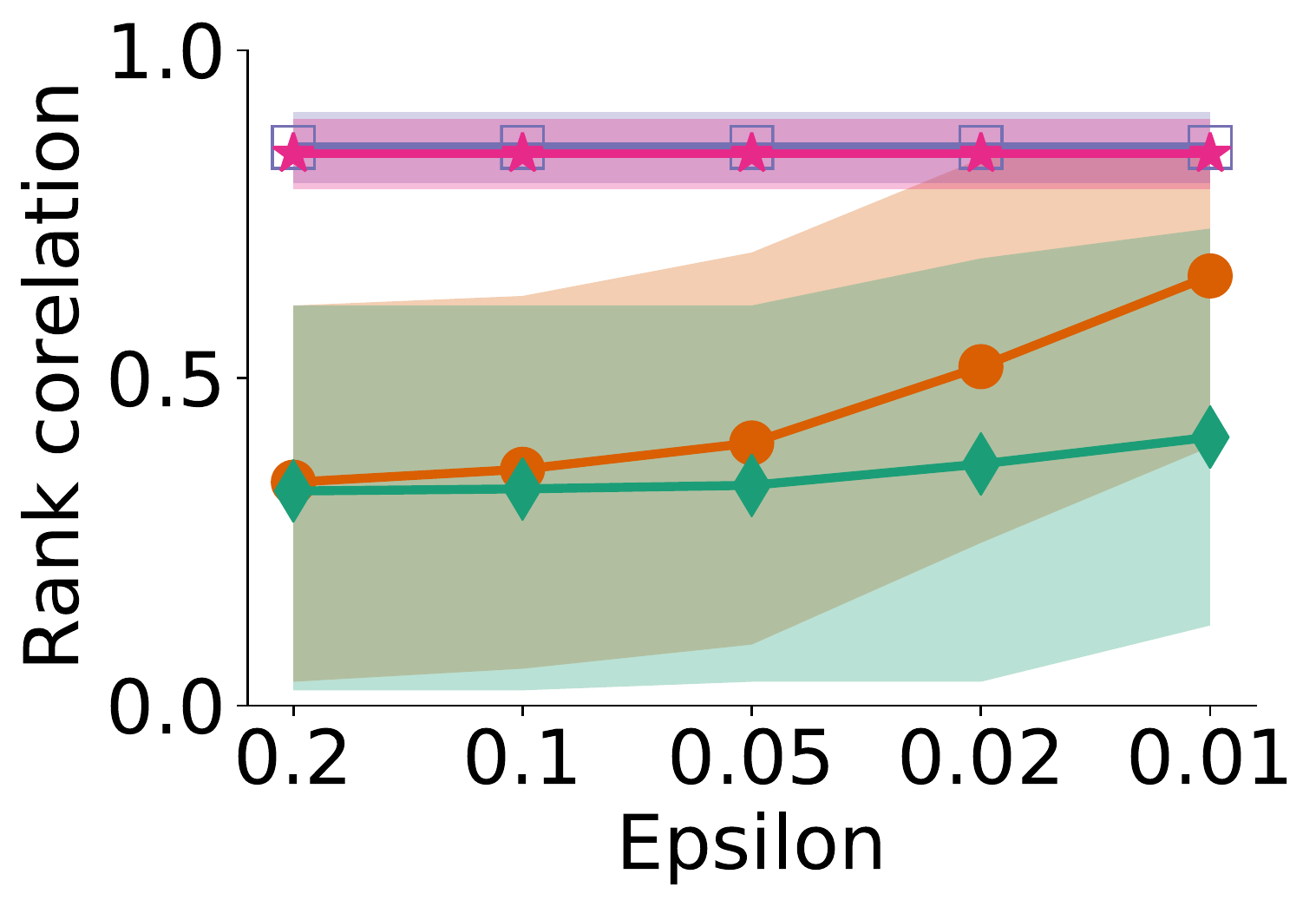}
		\caption{Flickr} 
	\end{subfigure}	
	\begin{subfigure}{0.18\textwidth}
		\includegraphics[width=\linewidth]{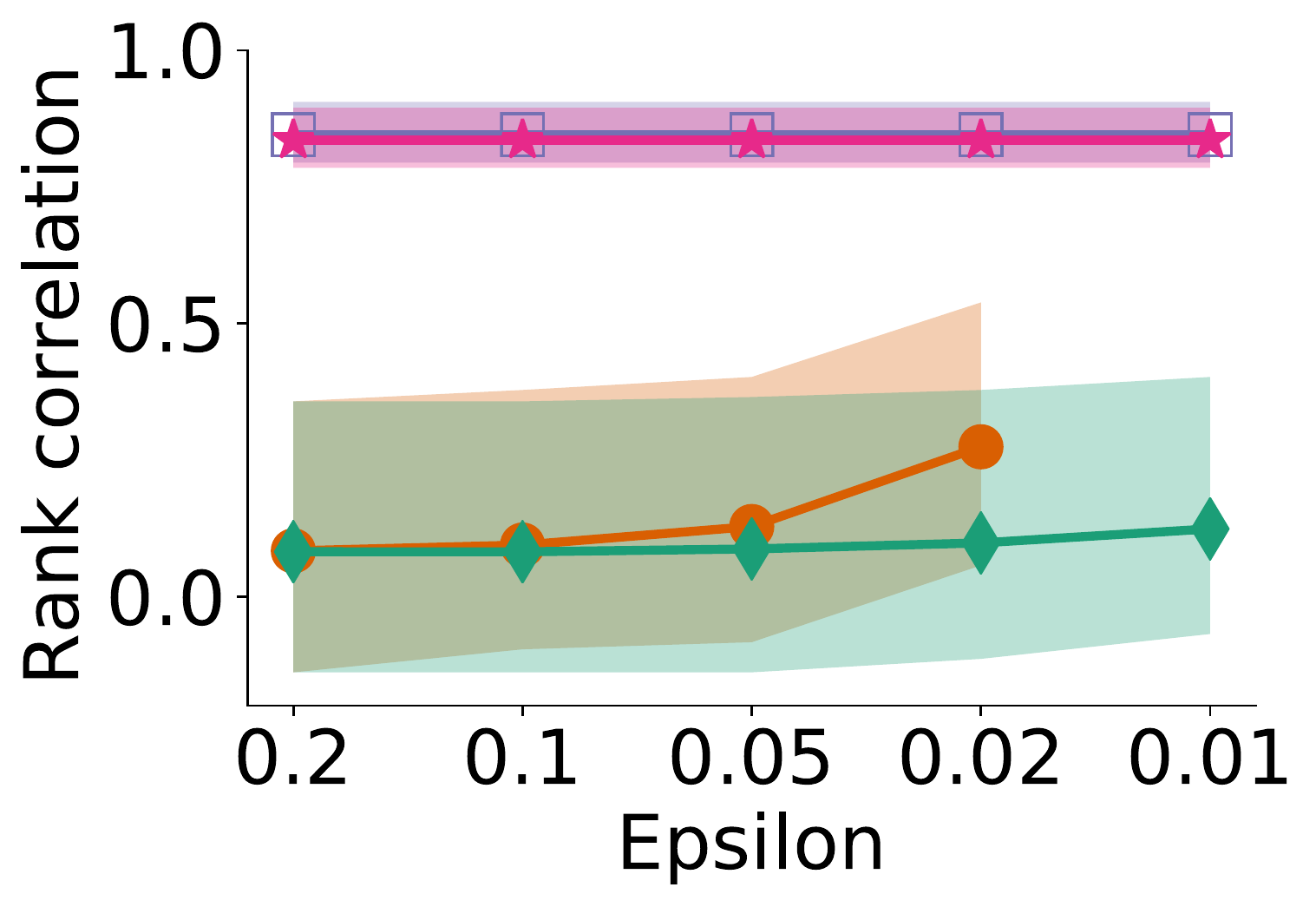}
		\caption{Livejournal} 
	\end{subfigure}	
	\begin{subfigure}{0.18\textwidth}
		\includegraphics[width=\linewidth]{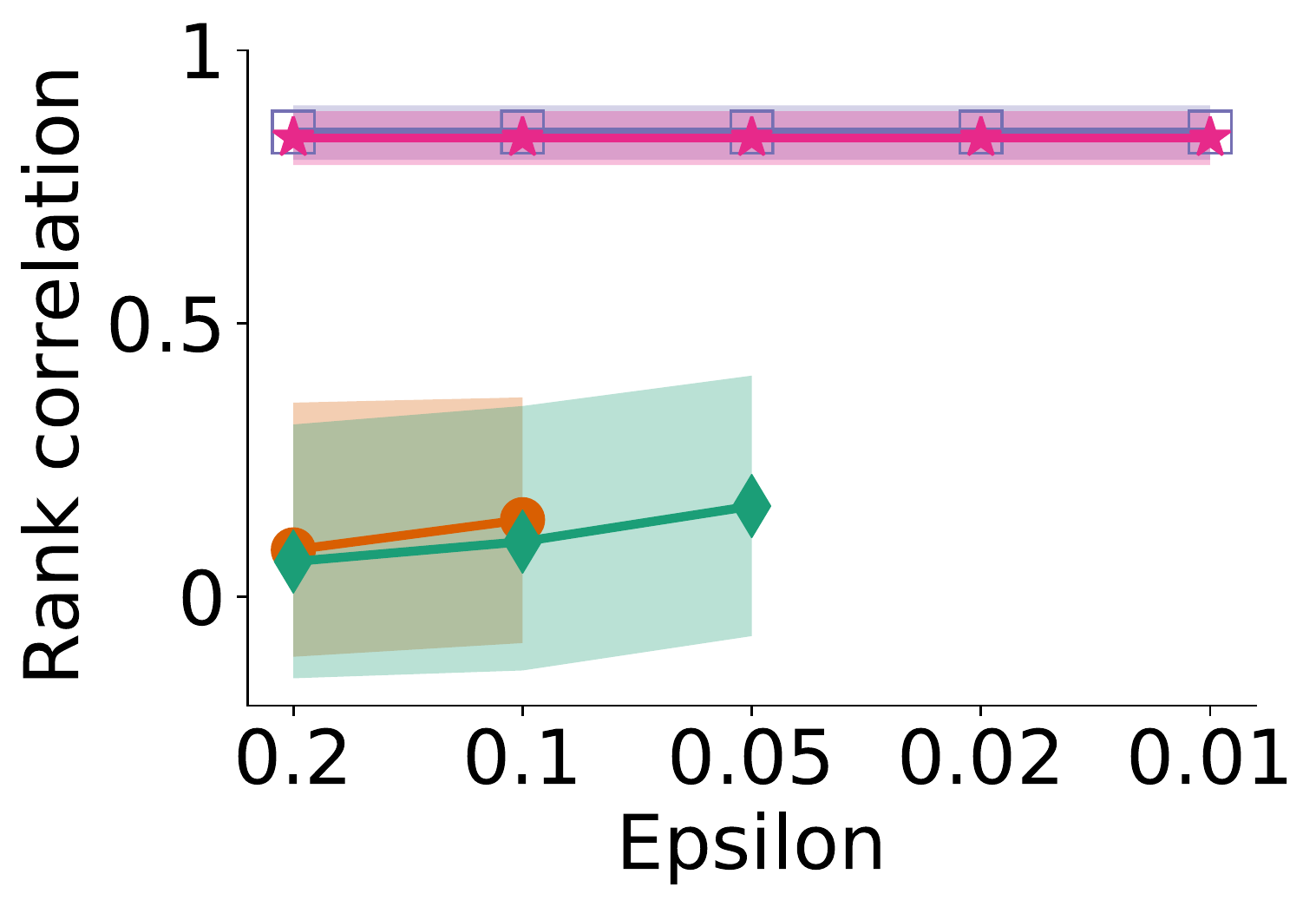}
		\caption{USA-road} 
	\end{subfigure}	
	\begin{subfigure}{0.18\textwidth}
		\includegraphics[width=\linewidth]{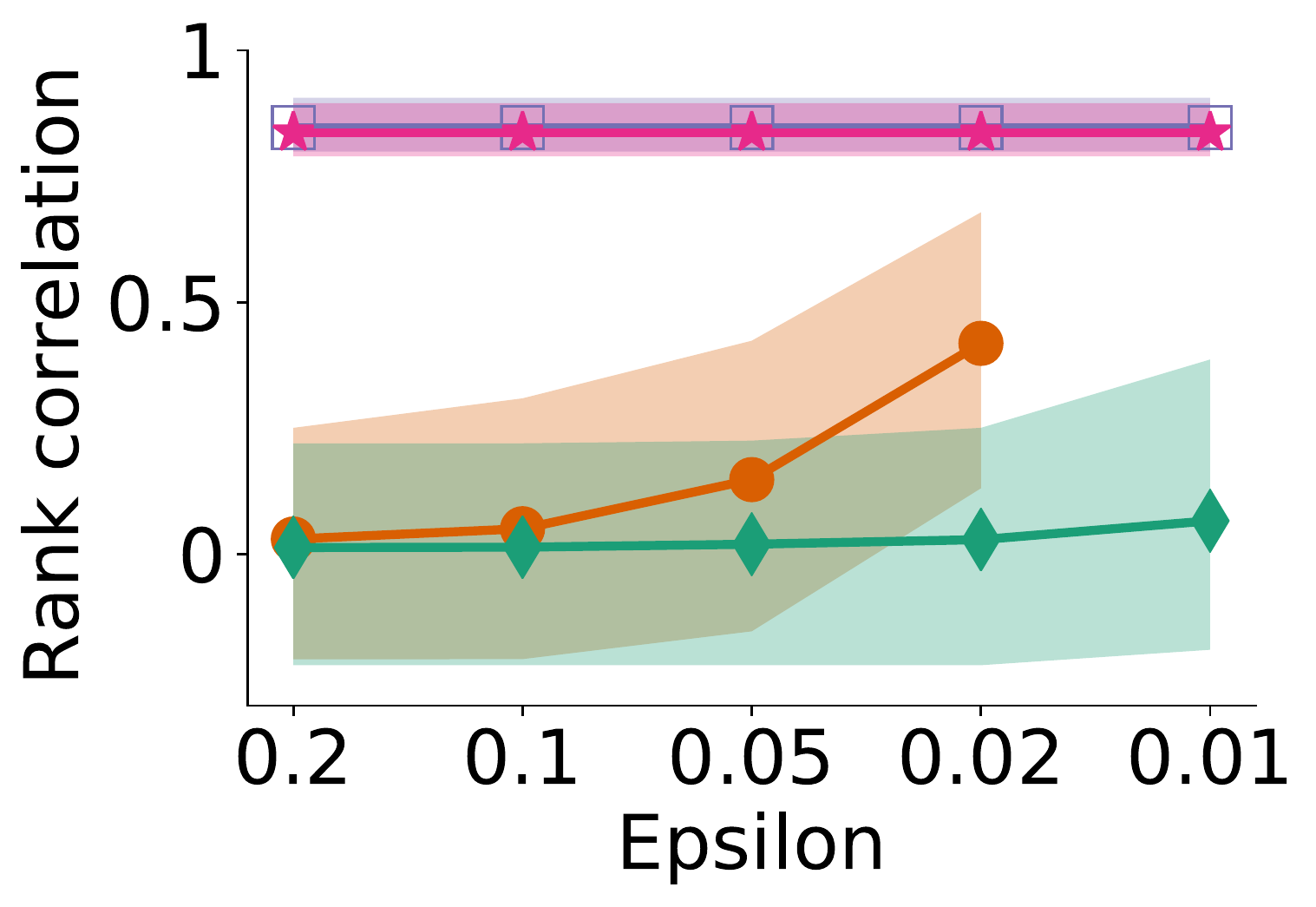}
		\caption{Orkut} 
	\end{subfigure}	
\vspace{-0.2cm}
	\caption{\small Rank correlation  at different error guarantees $\epsilon$. Shading areas show the $95\%$ confident intervals.}
	\label{fig:rankquality}
	\vspace{-0.6cm}
\end{figure*}

\subsection{Experiments settings}

\noindent\textbf{Algorithms.} We compare \RP{} algorithm, that is described in Subsection \ref{subsec:rp}, with \ABRA{} \cite{Riondato16kdd} (that uses node-pair sampling) and \KADBRA{} \cite{Borassi16} (that uses path sampling with bi-directed BFS). Note that, both \ABRA{} and \KADBRA{} can only estimate the betweenness centrality for the whole network. 
We also show the experiment result on \RP-full, i.e., the \RP{} algorithm with the subset of nodes is the whole network.


\vspace{-0.2cm}
\begin{table}[!h] \small
	\centering
	\caption{\small Networks' summary. }
		\vspace{-0.2cm}
	\begin{tabular}{ | l | r | r | r |}
		\hline
		\multirow{ 1}{*}{\bf Networks} & \multirow{ 1}{*}{\bf \#Nodes}  & \multirow{ 1}{*}{\bf \#Edges}  & \multirow{ 1}{*}{\bf Diam.} \\ 		
		\hline     
		\textbf{Flickr} &1.6 M&15.5  M&24   \\ \hline        
		
		\textbf{LiveJournal} &5.2 M&49.2 M&23   \\ \hline
		
		\textbf{USA-road} &23.9 M&58.3 M&1524  \\ \hline
		\textbf{Orkut} &3.1 M&117.2 M &10 \\ \hline
	\end{tabular}
	\label{tab:networks}
	\vspace{-0.1in}
\end{table}
\noindent\textbf{Networks and subsets.} We use 4 real-world networks from \cite{SNAP,DIMACS} as shown in Table \ref{tab:networks}. We ignore the information on the weight and direction of the edges, treating the networks as undirected and unweighted.  Unless  otherwise mentioned, in our experiments, we select $1000$ different subsets in which each subset consists of $100$ random nodes. We set $\epsilon$ to $0.05$ and $\delta$ to $0.01$.

\noindent\textbf{Ground truth.} We use the ground truth for Flickr, LiveJournal, and Orkut provided in  \cite{AlGhamdi17}. The ground truth was found in \cite{AlGhamdi17} by running  a parallel version of the Brandes algorithm  on a Cray XC40 supercomputer with 96,000 CPU cores and roughly 400TB of RAM. It took 2 million core hours (or roughly 10 years of calculations) to complete the calculation for 20 networks \cite{AlGhamdi17}. We obtain the ground truth for USA-road network using a parallel version of the Brandes's algorithm on our server with 96 Xeon E7-8894 CPUs (and 6TB memory) in about 2 weeks.  

\noindent\textbf{Metrics.} We compare the performance of the algorithms based on the following metrics. 

\begin{itemize}
	\item \emph{Running time.} Here, we exclude the time to load the network when we measure the running time.
	\item \emph{Rank quality.} For rank quality, we compute the \emph{Spearman's rank correlation} (see Eq.\ref{eq:spearman}) between the estimation 
	and the ground truth. 
	Note that, when we compute the rank of nodes, if there are two nodes with the same betweenness centrality, we break the tie by the nodes' IDs. 
	\item \emph{(Signed) relative error.} For a node $v$, let $bc(v)$ be the betweenness centrality of $v$ and $\tilde{bc}(v)$ be the estimation, the  relative error is given as $\left(\frac{\tilde{bc}(v)}{{bc}(v)} -1 \right) \times 100 \%.$
	In the case where ${bc}(v) = 0$, if  $\tilde{bc}(v) = 0$, the relative error is $0$. Otherwise, 
	the relative error is $\infty$.
\end{itemize}

    \begin{figure*}[!ht]
    	\centering 
    	\begin{subfigure}{0.18\textwidth}
    		\includegraphics[width=\linewidth]{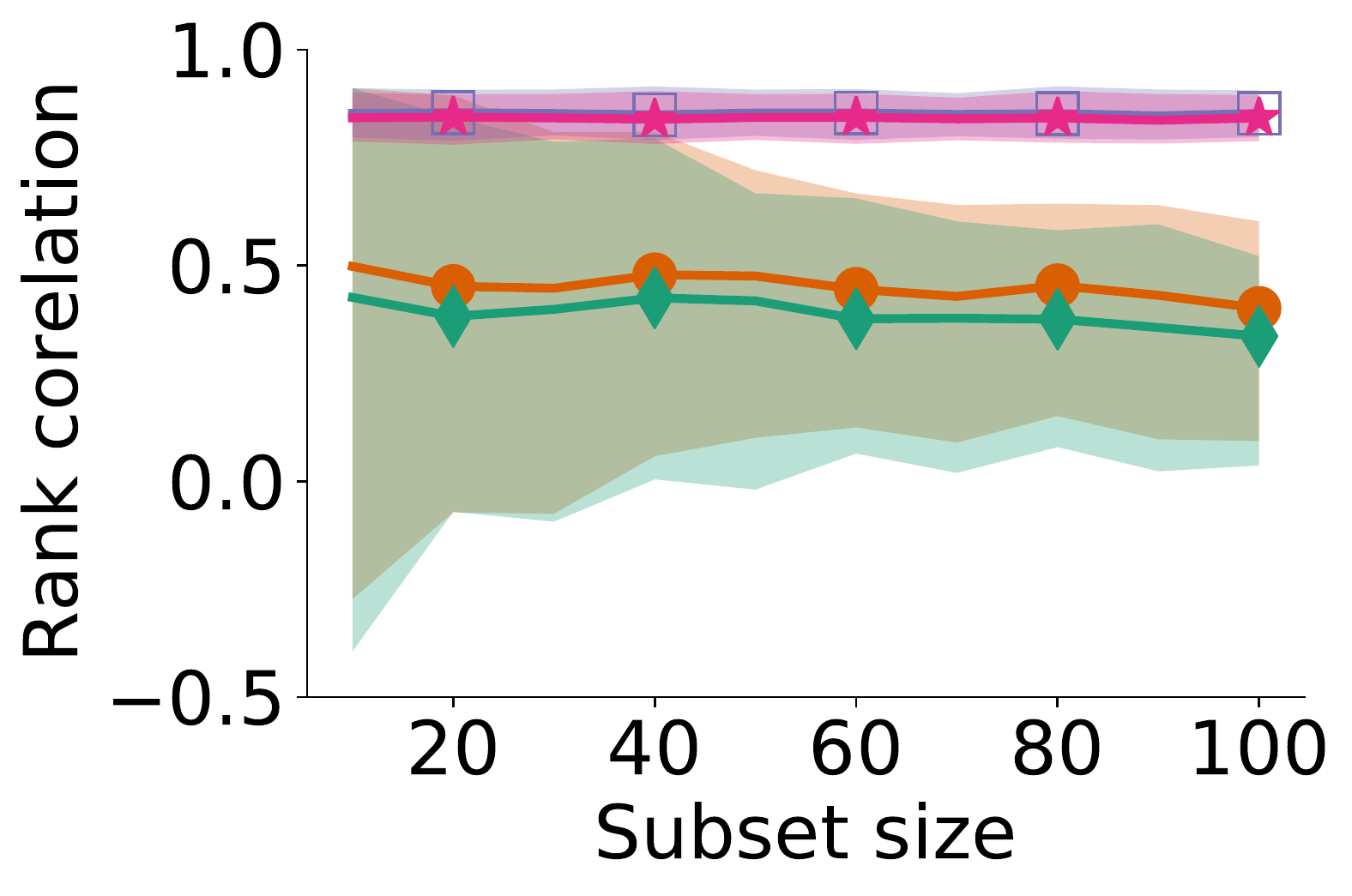}
    		\caption{ Flickr \\ \phantom{a}} 
    	\end{subfigure}	
    	\begin{subfigure}{0.18\textwidth}
    		\includegraphics[width=\linewidth]{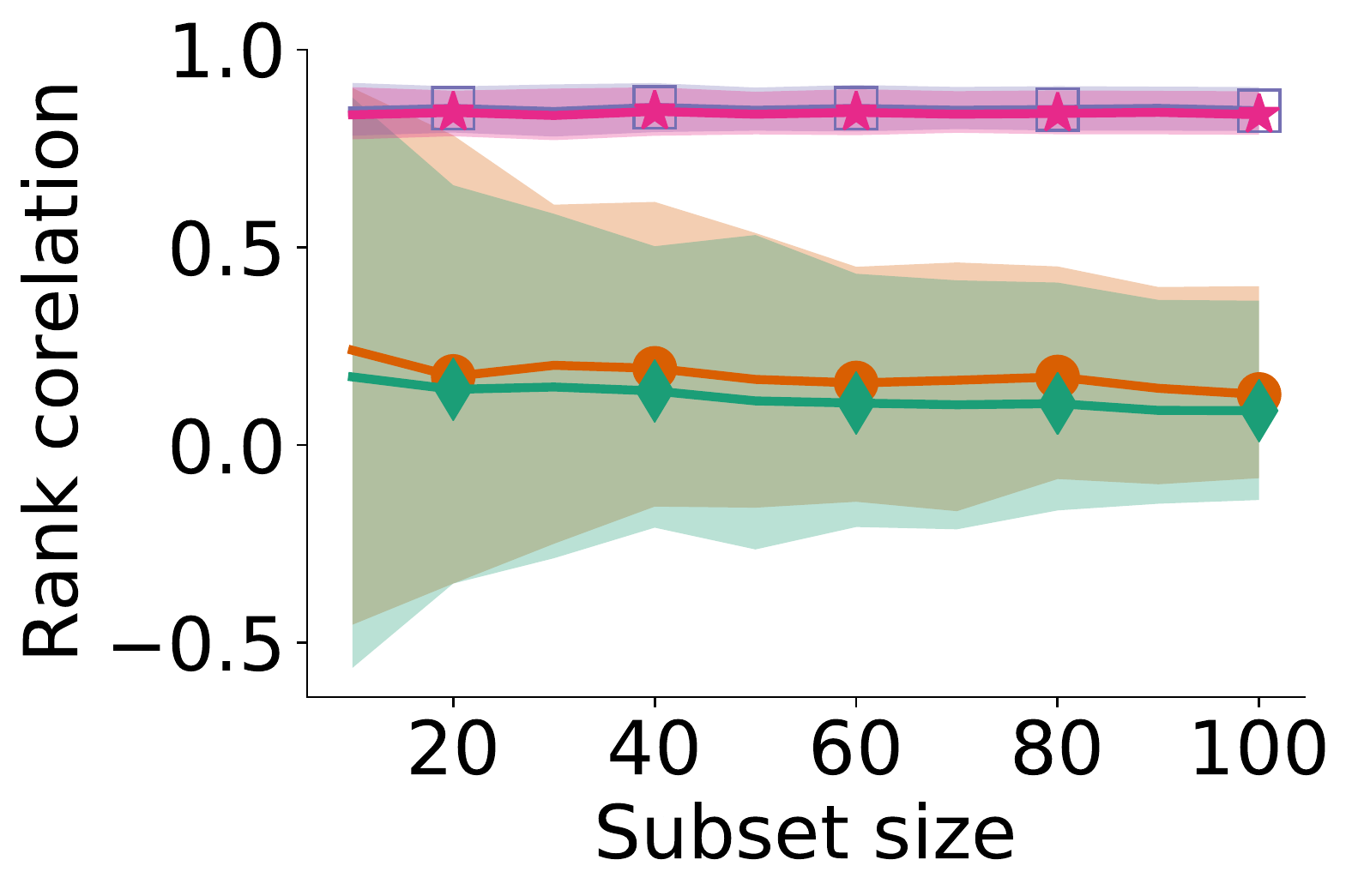}
    		\caption{ LiveJournal  \\ \phantom{a}} 
    	\end{subfigure}	
    	\begin{subfigure}{0.18\textwidth}
    		\includegraphics[width=\linewidth]{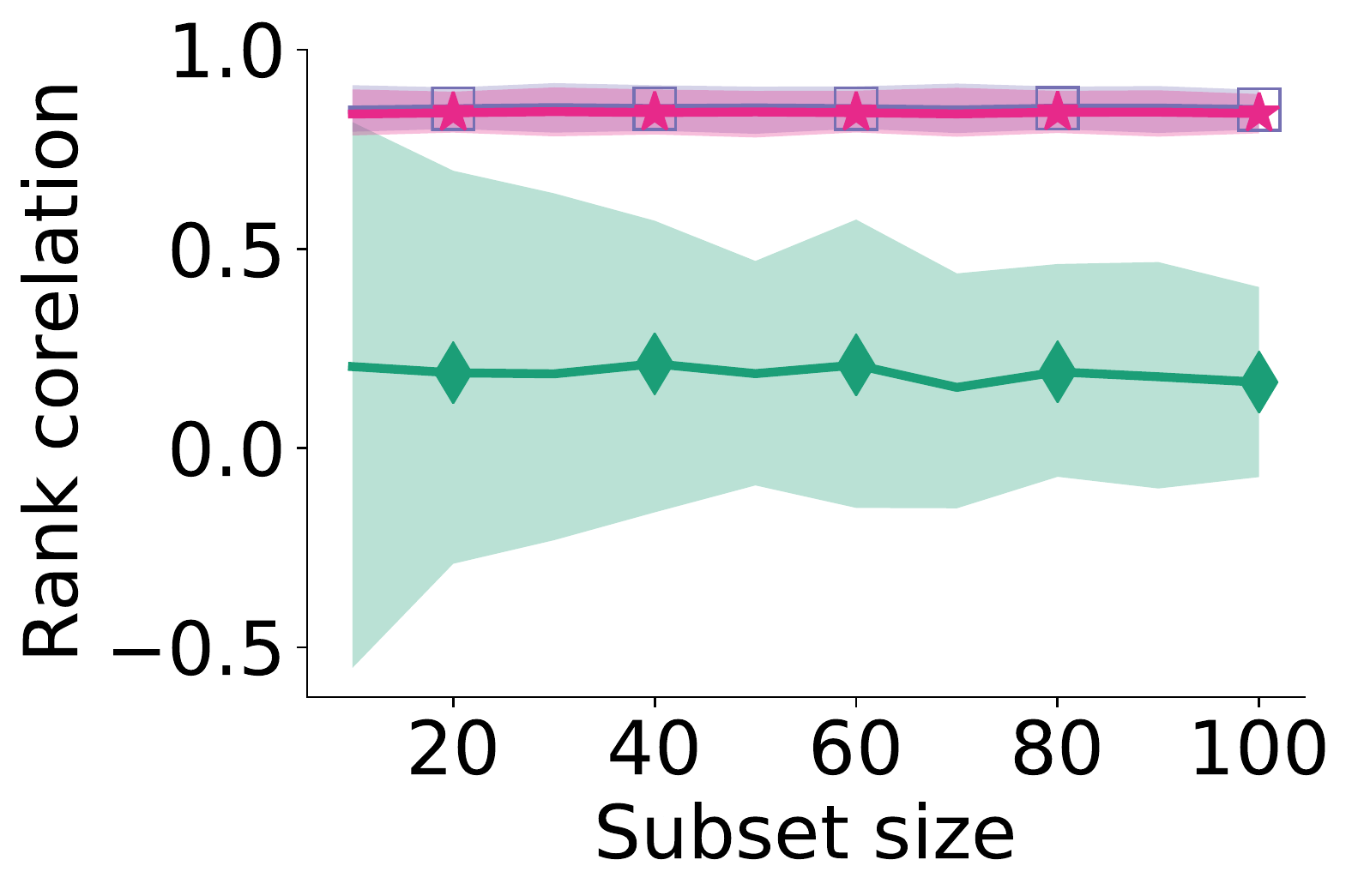}
    		\caption{ USA-road (\ABRA{} cannot finish in $10$ hours)} 
    	\end{subfigure}	
    	\begin{subfigure}{0.18\textwidth}
    		\includegraphics[width=\linewidth]{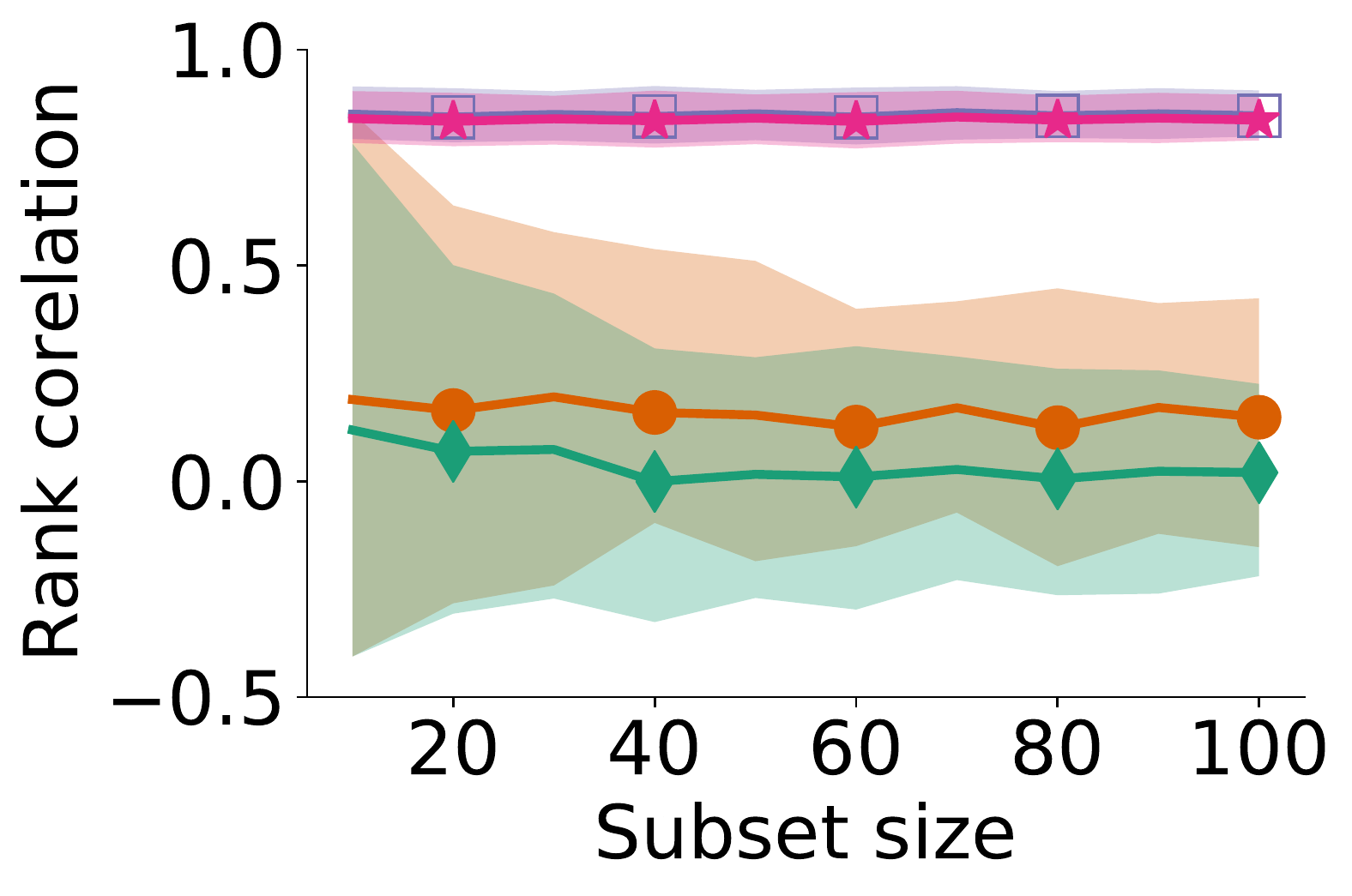}
    		\caption{ Orkut \\ \phantom{a}} 
    	\end{subfigure}	
    \vspace{-0.2cm}
    	\caption{\small Rank correlation with different subset sizes. The shaded areas show the $95\%$ confident intervals.}
    	\label{fig:rankqualitydis}
    	\vspace{-0.5cm}
    \end{figure*}

 

\noindent\textbf{Environment.} We implemented our algorithms in C++ and obtained the implementations of others from the corresponding authors. We conducted all experiments on a CentOS machine Intel(R) Xeon(R) CPU E7-8894 v4 2.40GHz.
We set the time limit to 10h (36,000s).

%
%
\subsection{Experiment results}
%

First, we run an experiment with varying $\epsilon \in \{0.2,0.1,0.05,0.02,0.01\}$ and $\delta = 0.01$. We select $1,000$ different subset in which each subset consists of $100$ random nodes.

\noindent\textbf{Running time.} 
From Fig. \ref{fig:runningtime}, the running time of \RP{} $7-235$ times smaller than \KADBRA{} and $90-425$ times faster than \ABRA.
In 10 hours, \ABRA{} can not finish  in $5$ cases and \KADBRA{} can not finish  in $2$ cases. 

Furthermore, the running time of \RP{} on a set of target nodes is also better than the running time of \RP{}-full. Indeed, on average \RP{} runs $4-11$ times faster than \RP{}-full.

\noindent{\textbf{Rank correlation.} }
\RP{} and \RP{}-full always provides a better ranking correlation, in comparison with \ABRA{} and \KADBRA{} (see Fig. \ref{fig:rankquality}). 
For example, in LiveJournal graph, for $\epsilon = 0.05$, the  Spearman's rank correlation of the estimation of \RP{} and the ground truth is $0.84$. The rank correlation of the estimation of \ABRA{} and \KADBRA{} are $0.13,0.09$, respectively. Furthermore, the rank quality of \ABRA{} and \KADBRA{}  are widely varying. For example, for  $\epsilon = 0.05$, the rank correlation of \ABRA{} varies from $0.12$ to $0.63$ and the rank correlation of \KADBRA{} varies from $0.02$ to $0.58$.

We also run an experiment with fixed $\epsilon = 0.05$ and varying subset size from $10$ to $100$. As shown in Fig. \ref{fig:rankqualitydis}, the varying range of the rank quality of \ABRA{} and \KADBRA{} increases as the subset size decreases.

    \begin{figure*}[!ht]
	\centering 
	\begin{subfigure}{0.18\textwidth}
		\includegraphics[width=\linewidth]{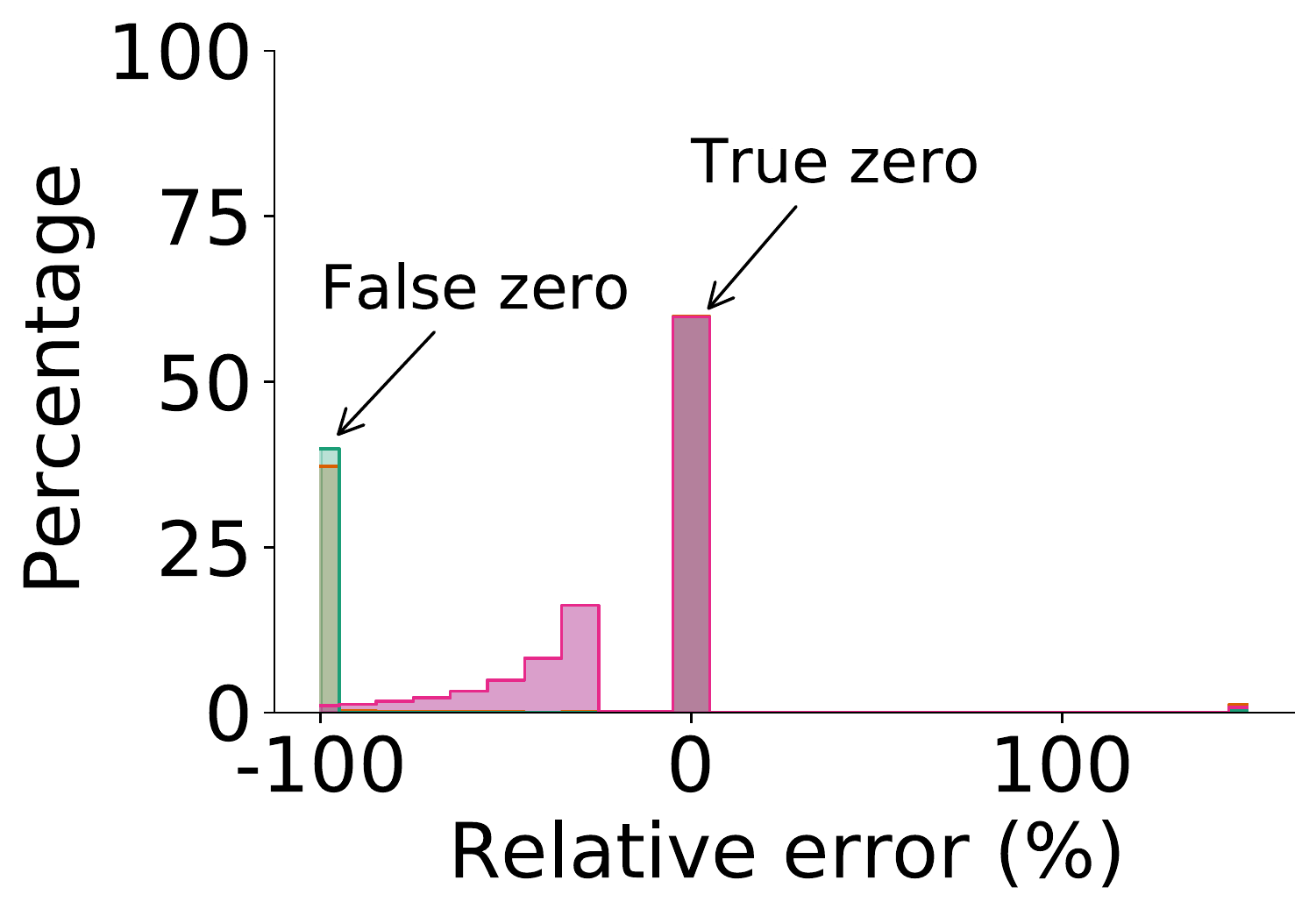}
		\caption{Flickr \\ \phantom{a}} 
	\end{subfigure}	
	\begin{subfigure}{0.18\textwidth}
		\includegraphics[width=\linewidth]{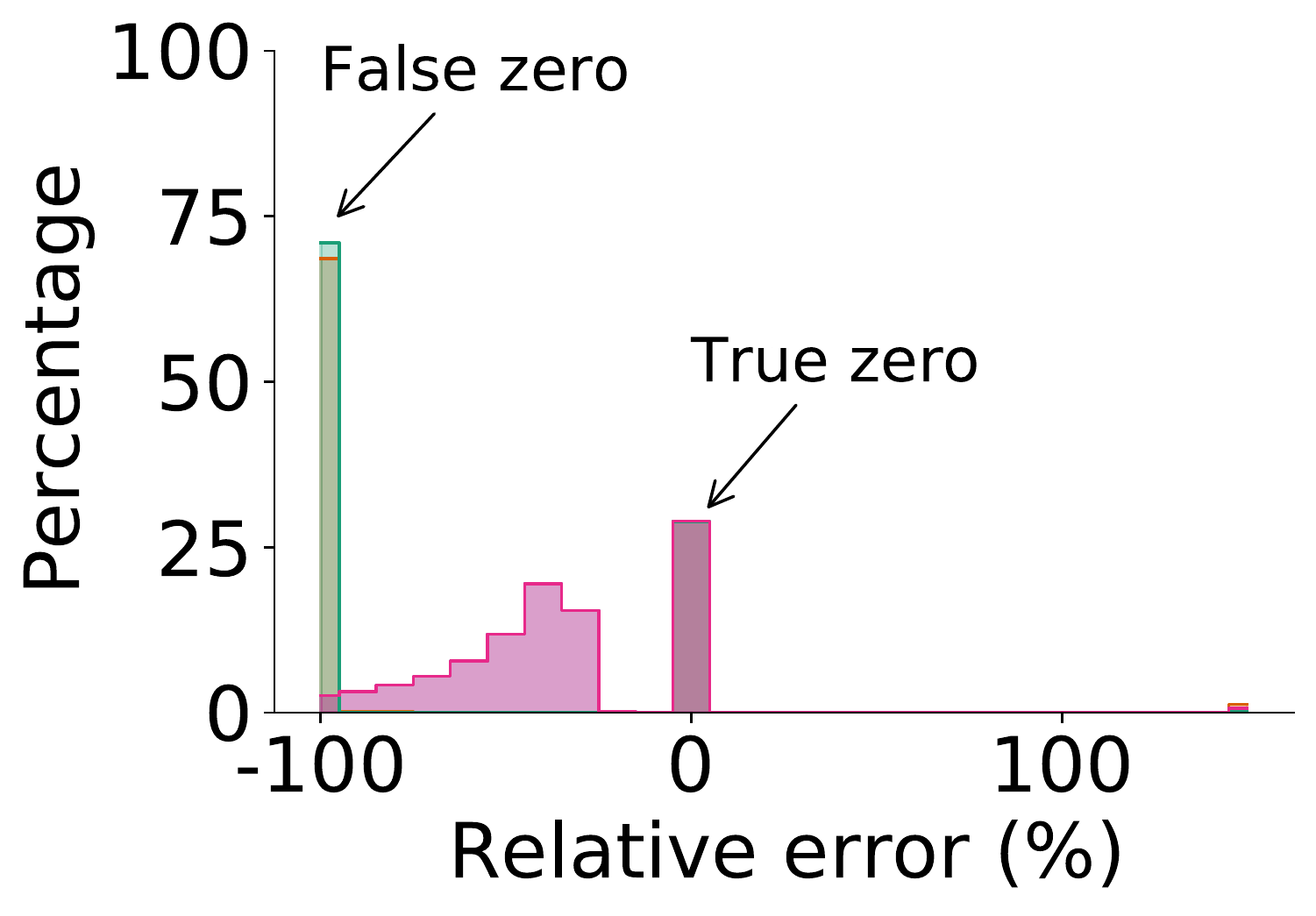}
		\caption{LiveJournal \\ \phantom{a}} 
	\end{subfigure}	
	\begin{subfigure}{0.18\textwidth}
		\includegraphics[width=\linewidth]{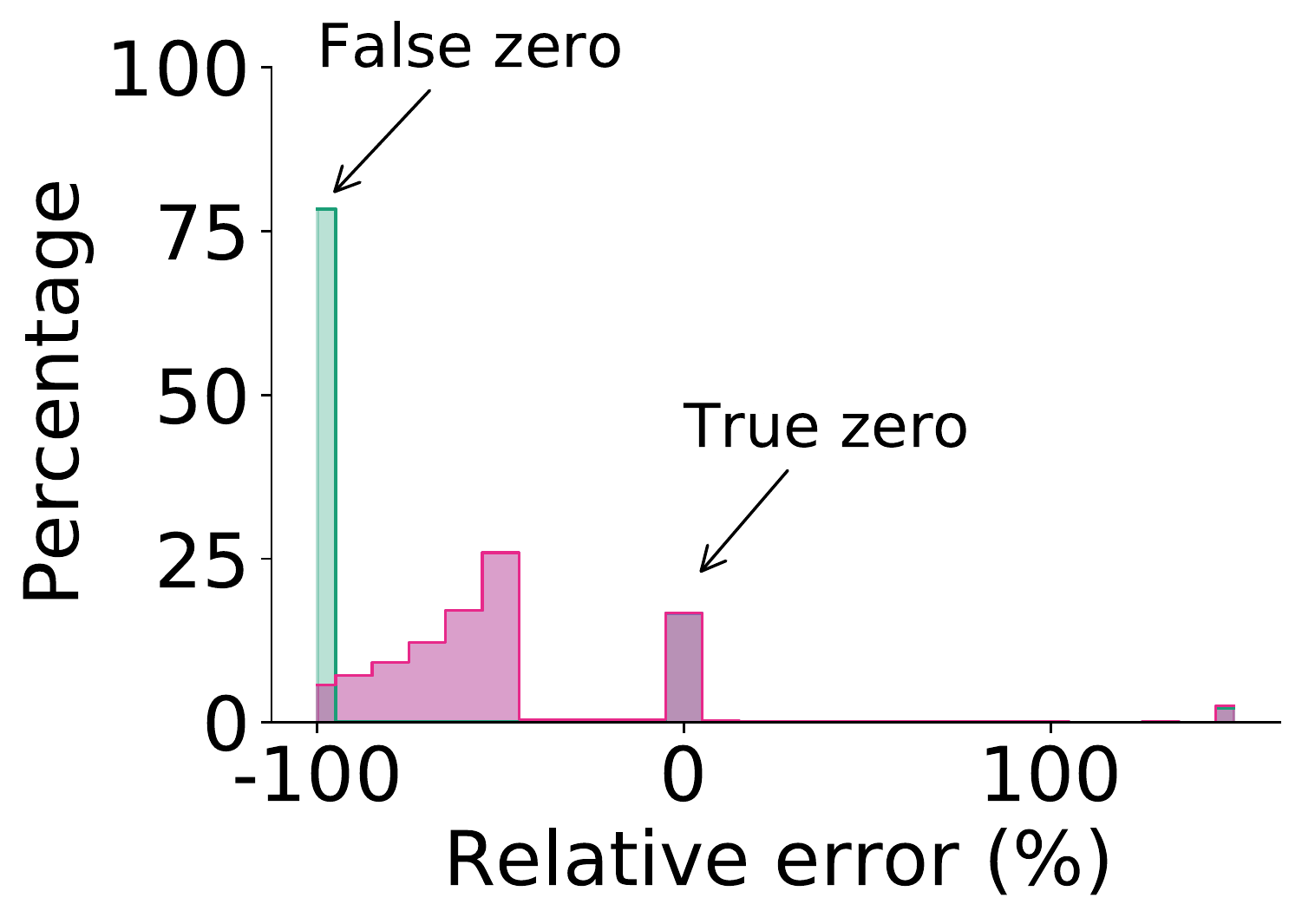}
		\caption{USA-road (\ABRA{} cannot finish in $10$ hours.)} 
	\end{subfigure}	
	\begin{subfigure}{0.18\textwidth}
		\includegraphics[width=\linewidth]{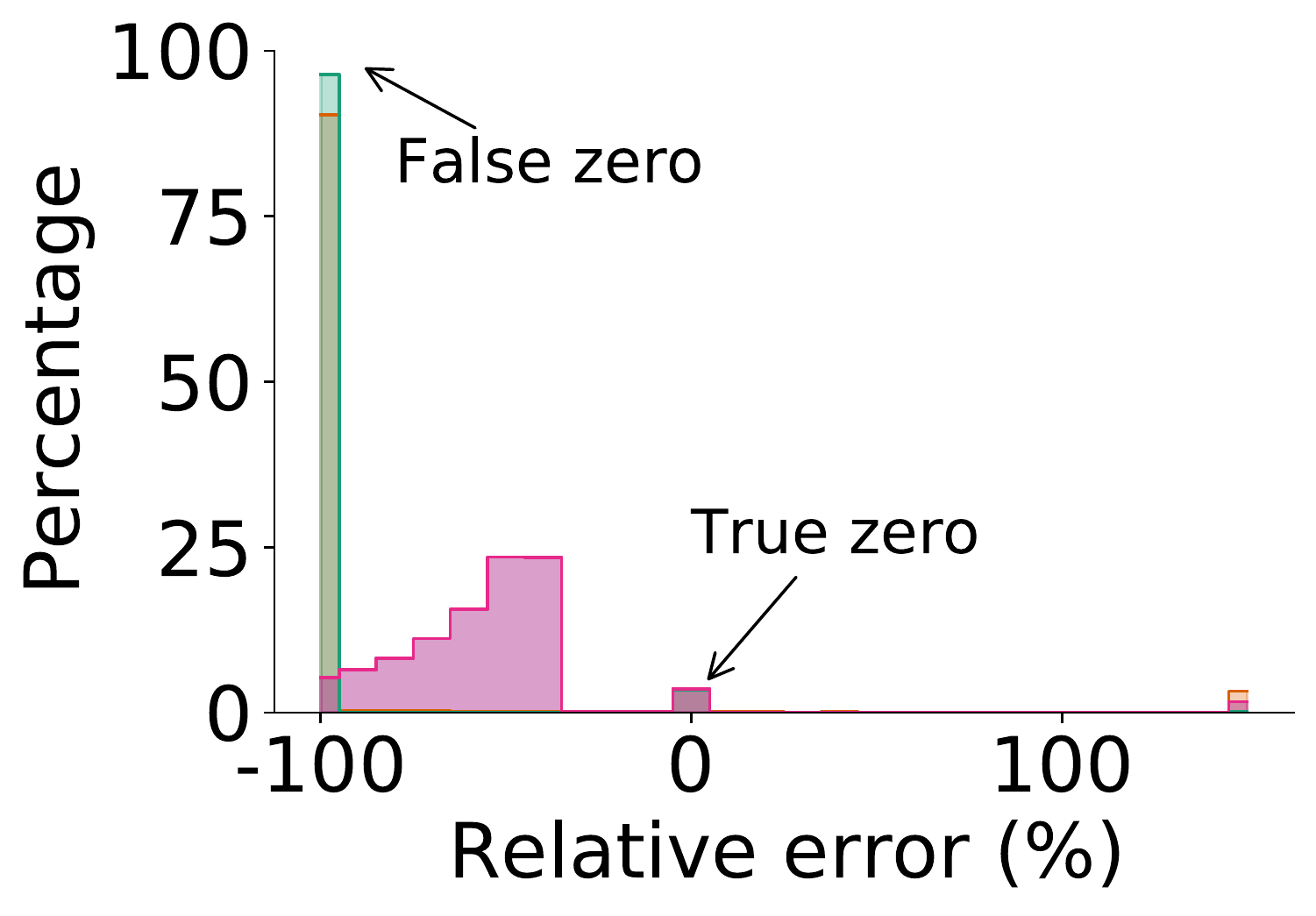}
		\caption{Orkut \\ \phantom{a}} 
	\end{subfigure}	
\vspace{-0.2cm}
	\caption{\small (Signed) relative error}
	\label{fig:Rerror}
	\vspace{-0.6cm}
\end{figure*}
\begin{figure*}[!ht]
	\centering 
	\begin{subfigure}{0.4\textwidth}
		\includegraphics[width=\linewidth]{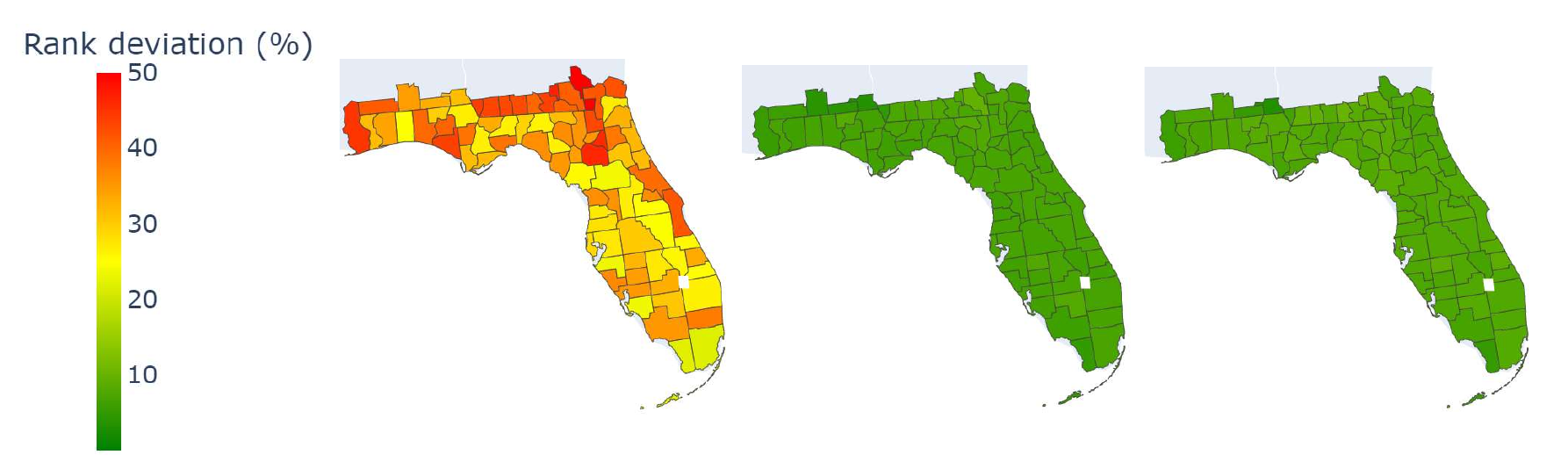}
		\caption{The rank deviation of \KADBRA, \RP-full, \RP{} (from left to right) on Florida. \ABRA{} cannot finish in $10$ hours.} 
		\label{fig:map}
	\end{subfigure}	
	\begin{subfigure}{0.24\textwidth}
		\includegraphics[width=\linewidth]{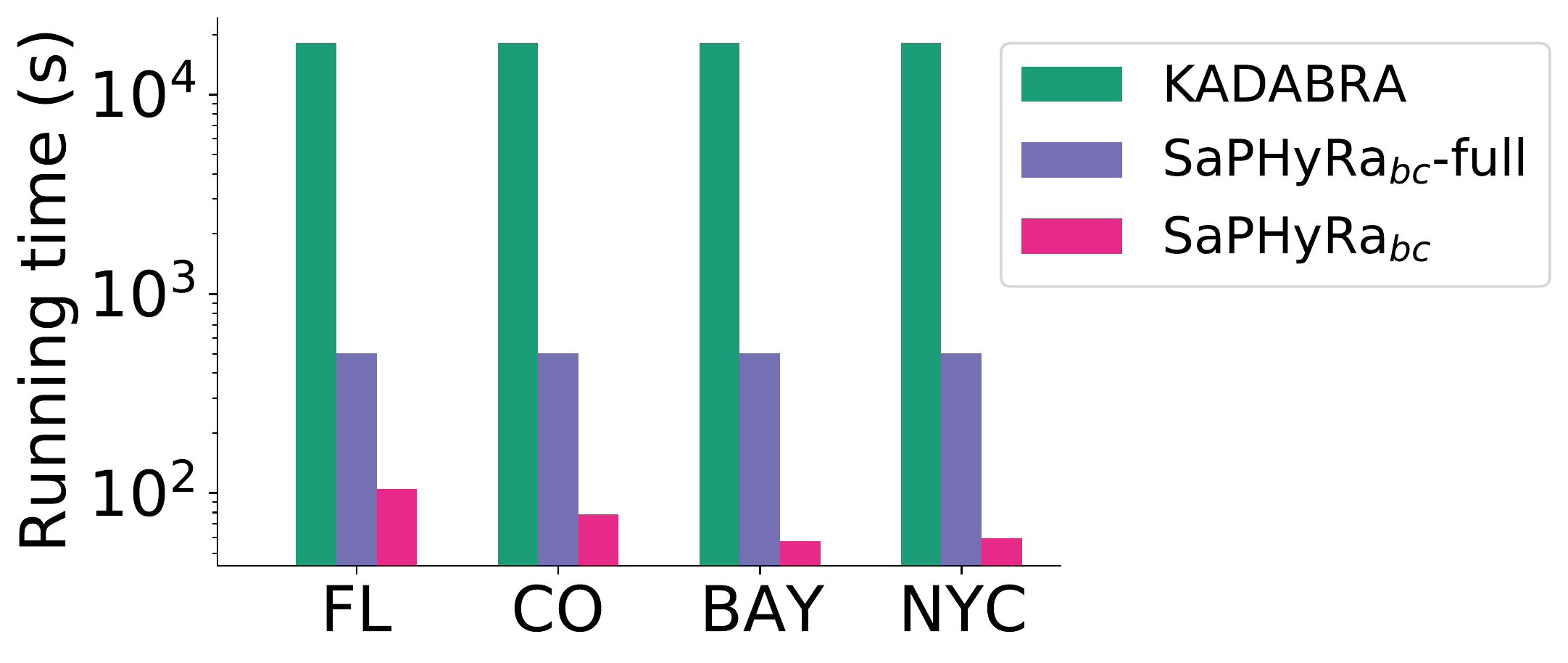}
		\caption{The running time} 
	\end{subfigure}	
	\begin{subfigure}{0.14\textwidth}
		\includegraphics[width=\linewidth]{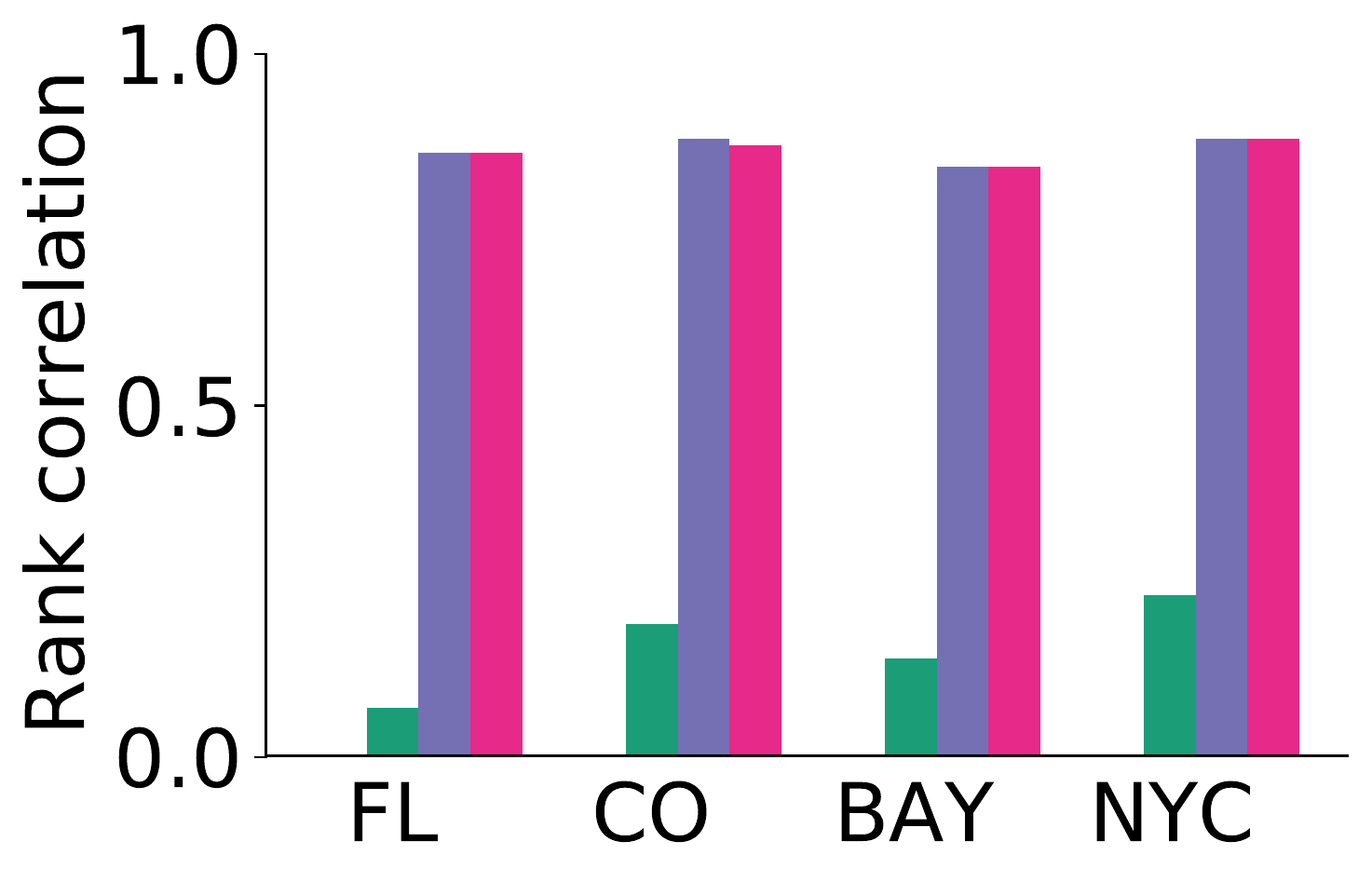}
		\caption{The rank quality} 
	\end{subfigure}	
\vspace{-0.2cm}
	\caption{USA-road}
	\label{fig:USA}
	\vspace{-0.6cm}
\end{figure*}
\noindent{\textbf{Relative error.}} We measure the relative error in an experiment where $\epsilon = 0.05$ and the subset size is $100$.
In Fig. \ref{fig:Rerror}, we show the histogram of the relative error of the estimations with the ground truth.  Here, we group all nodes that have relative error bigger than $150\%$ to a single bucket. 

From Fig. \ref{fig:Rerror}, we observe that for \ABRA and \KADBRA, more $95\%$ of nodes that have the relative error equal either $0$ or $-100\%$.
A close investigation reveal that those are the nodes with an estimated centrality zero. Those can be further divided into \emph{true zeros}: nodes with betweeness centrality that are correctly estimated as zeros and \emph{false zeros}: nodes with positive betweenness centrality that are incorrectly estimated as zeros.

Combine with the rank quality in Fig. \ref{fig:rankquality}, we have the following observations. 
\begin{itemize}
	\item \emph{The more true zeros, the higher rank quality.} From Fig. \ref{fig:Rerror}, the fractions of true zeros are $59,29,16,4 \%$ on Flickr, LiveJournal, USA-road, Orkut, respectively. For a node $v$ with ${bc}(v) = 0$, all the studied algorithms will return $0$ as the estimation, i.e., true zeros are the easy cases that cannot be incorrectly estimated. Since the true zero in Flickr network is higher, \ABRA{} and \KADBRA{} provide the estimation with better rank quality on Flickr (see Fig. \ref{fig:rankqualitydis}). 
	\item \emph{The fewer false zeros, the higher rank quality.} From Fig. \ref{fig:Rerror}, the fractions of false zeros for \ABRA{} are $37,68,90 \%$ on Flickr, LiveJournal, Orkut, respectively. For \KADBRA{}, the percentages are $39,71,78,96\%$ on Flickr, LiveJournal, USA-road, Orkut, respectively. For \RP-full and \RP{}, as we have shown in Lemma \ref{lemma:falsezero}, there will be no false zeros.  As a result, \RP-full and \RP{} can provide the estimation with better rank quality than \ABRA{} and \KADBRA{}. 
\end{itemize}

\vspace{-0.1cm}
\noindent {\textbf{Case study on USA-road.}} Here, we select $4$ subsets as $4$ areas in \cite{DIMACS} (see  Table \ref{tab:subset} for summary). More concretely, 
we extract the longitude, latitude of nodes in $4$ areas, and map them with the node in USA-road network. 
\vspace{-0.2cm}
\begin{table}[!h] \small
	\centering
	\caption{\small Subset' summary. }
	\begin{tabular}{ | l | r | r |}
		\hline
		\multirow{ 1}{*}{\bf Networks} & \multirow{ 1}{*}{\bf \#Nodes}  & \multirow{ 1}{*}{\bf \#Edges}  \\ 		
		\hline     
		\textbf{New York City (NYC)} &264 K&734 K  \\ \hline        
		
		\textbf{San Francisco Bay Area (BAY)} &321 K&800 K \\ \hline
		
		\textbf{Colorado (CO)} &435 K& 1,057 K  \\ \hline
		\textbf{Florida(FL)} &1,070 K&2,713 K \\ \hline
	\end{tabular}
	\label{tab:subset}
	\vspace{-0.15in}
\end{table}

Similar to the previous experiments, \RP-full and \RP{} outperform \KADBRA{} on both running time and rank quality (Fig. \ref{fig:USA} ). 
Furthermore, the running time of \RP{} is better with the subset size smaller size. For example, as the subset size reduces from $1,070 K$ (FL) to $264 K$ (NYC), 
the running time of \RP{} reduces from $105s$ to $59.4s$. 

In Fig. \ref{fig:map}, we show the average rank deviation  of nodes in the areas of Colorado. 
\RP-full and \RP{} outperforms \KADBRA{} in term of rank deviation (\ABRA{} cannot finish in 10 hours). For  \KADBRA{}, the highest average rank deviation in an area is $39\%$. Meanwhile, the highest average rank deviation in an area of \RP-full and \RP{} are $11\%,12\%$, respectively.

\section{Conclusion.}
 We propose and investigate the  ranking subset problem  when it is computationally prohibitive to obtain the exact centrality values. Our proposed  \SSP{} framework indicates the possibility to reduce  the running time significantly when ranking a subset in contrast to ranking all nodes in the network. It also demonstrates an effective way to hybrid good estimation heuristics with sampling-based estimation methods to obtain both high ranking quality and theoretical guarantees on the error. Future directions include extending the framework to other centrality measures such as 
closeness centrality, nodes' influence, and Shapley value. Further, designing ranking methods with \emph{provable guarantees on the ranking} (not just the estimation errors) is of particular interest.